\documentclass[3p]{elsarticle}

\usepackage{a4wide}
\usepackage{float}
\usepackage{amsmath}
\usepackage{amssymb}
\usepackage{graphicx}
\usepackage{caption}
\usepackage{subcaption}
\usepackage[latin1]{inputenc} 
\usepackage{rotating}
\usepackage{setspace}
\usepackage{sidecap}
\usepackage{upgreek}
\usepackage{microtype}
\usepackage{url}
\usepackage{lineno}
\begin{document}
\begin{frontmatter}

\pagenumbering{Roman}
\let\clearpage\relax

 \title{Impact of low-dose electron irradiation on n$^+$p silicon strip sensors}

 \renewcommand{\thefootnote}{\fnsymbol{footnote}}

   \address{\large\rm{(The Tracker Group of the CMS Collaboration\footnote{Corresponding author:~Robert Klanner, Hamburg University, Germany,
    \it{Email address}\rm{:~Robert.Klanner@desy.de}})}\normalsize}



 \begin{abstract}

The response of $n^+p$ silicon strip sensors to electrons from a $^{90}$Sr source was measured using a multi-channel read-out system with 25\,ns sampling time.
 The measurements were performed over a period of several weeks, during which the operating conditions were varied. The sensors were fabricated by Hamamatsu Photonics\,K.\,K. on 200\,$\upmu $m thick float-zone and magnetic-Czochralski silicon.
 Their pitch was 80\,$\upmu $m, and both $p$-stop and $p$-spray isolation of the $n^+$\,strips were studied.
 The electrons from the $^{90}$Sr source were collimated to a spot with a full-width-at-half-maximum of 2\,mm at the sensor surface, and the dose rate in the SiO$_2$ at the maximum was about 50\,Gy/d.
 After only a few hours of making measurements, significant changes in charge collection and charge sharing were observed.
 Annealing studies, with temperatures up to $80^\circ $C and annealing times of 18\,hours, showed that the changes can only be partially annealed.
 The observations can be qualitatively explained by the increase of the positive oxide-charge density due to the ionization of the SiO$_2$ by the radiation from the $\beta $\,source.
 TCAD simulations of the electric field in the sensor for different oxide-charge densities and different boundary conditions at the sensor surface support this explanation.
 The relevance of the measurements for the design of $p^+n$ strip sensors is discussed.

 \end{abstract}

 \end{frontmatter}
 \tableofcontents
 \newpage
 \pagenumbering{arabic}

 \section{Introduction}
   \label{sect:Introduction}

 Today, segmented silicon detectors with a spatial resolution of approximately $10\,\upmu$m  are used in precision tracking detectors closest to the interaction point of most collider experiments.
 They contribute to practically all physics analyses and were essential for the discovery of the Higgs boson and many other important physics results from the four large-scale CERN-LHC experiments.
 They have demonstrated an extraordinary performance with respect to precision, efficiency and reliability.
 The High-Luminosity LHC upgrade (HL-LHC) poses further challenges with respect to track density and radiation exposure:
 For an anticipated integrated luminosity of 3\,000\,fb$^{-1}$, hadron fluences of up to $10^{16}$\,n$_{eq}$/cm$^{2}$ (1\,MeV equivalent neutrons/cm$^{2}$) at a distance of about 5\,cm from the beam are expected from simulations, causing radiation damage to the silicon crystal and ionization doses of order 1\,MGy, resulting in surface damage to the insulating layers of the sensors.
 Whilst silicon-bulk damage has been studied extensively, only limited knowledge of surface  damage to high-ohmic silicon\,\cite{Zhang:2012} and its interplay with bulk damage is available.

 In this paper the effects of ionizing radiation for dose values between 0 and 1\,kGy on the charge collection properties of $n^+p$ strip sensors are studied using electrons from a 100\,MBq $^{90}$Sr source.
 The relevance of the results for the HL-LHC upgrade is discussed.
 More details about the measurements and the results can be found in Refs.\,\cite{Erfle:2014, Henkel:2014}.
 Some of the results have been presented at TIPP\,2014\,\cite{TIPP2014}.

   \section{Experimental setup and sensors investigated}
    \label{sect:Setup}

\begin{figure}[!ht]
   \centering
   \begin{subfigure}[a]{10cm}
    \includegraphics[width=10cm]{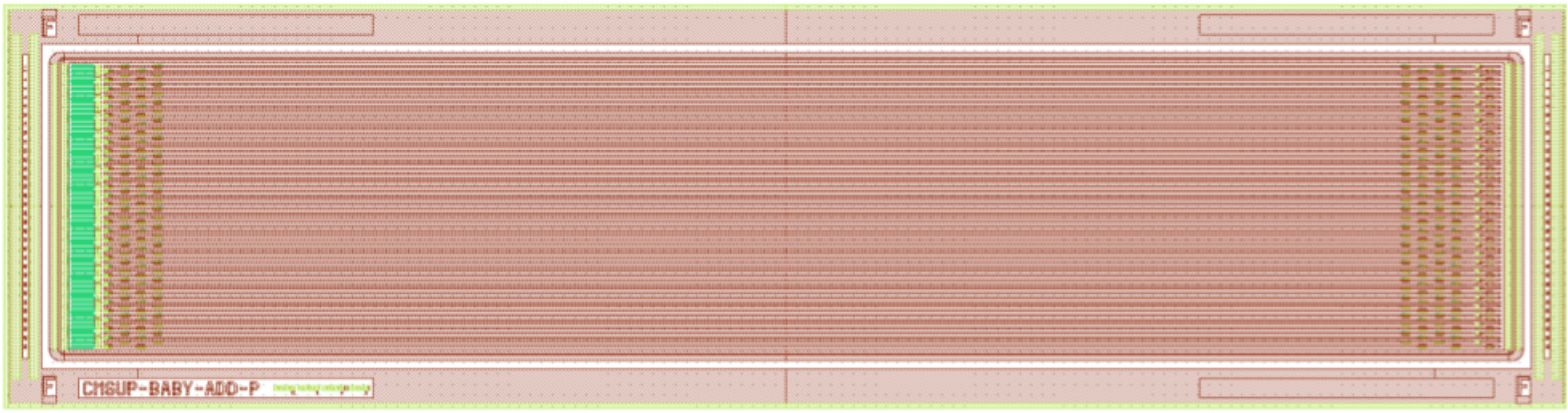}
    \caption{ }
     \label{fig:Sensor1}
   \end{subfigure}%
    ~
   \newline
    \begin{subfigure}[a]{12.5cm}
    \includegraphics[width=11cm]{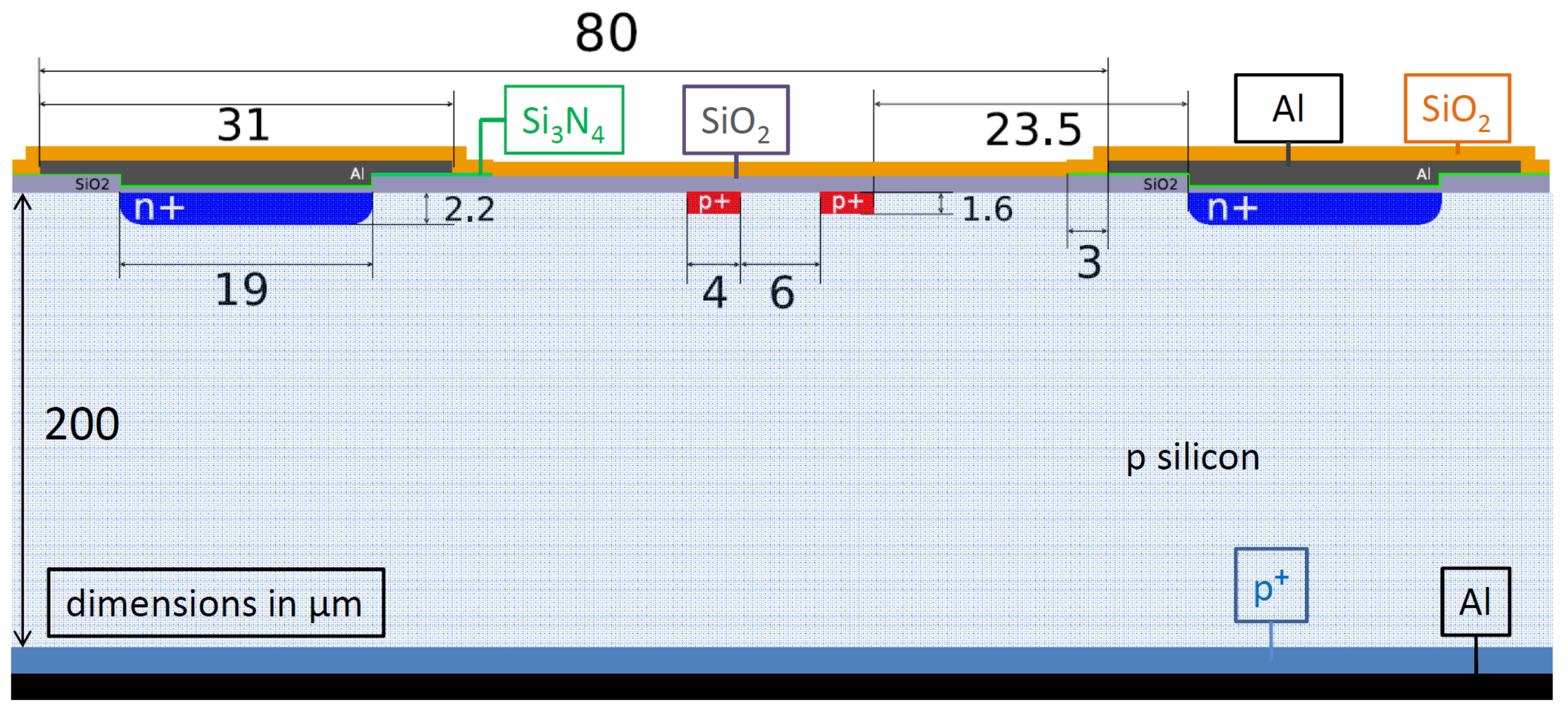}
    \caption{ }
     \label{fig:Sensor2}
   \end{subfigure}%
   \caption{\,Layout of the $p$-stop sensor. (a)\,Top view and (b)\,cross section. The numbers given, in particular the implantation depths, are only approximate. They have been used for the simulations discussed in Sect.\,\ref{sect:Discussion}.}
  \label{fig:Sensor}
\end{figure}

 Two different types of mini strip $n^+p$ sensors from the CMS-HPK Campaign\,\cite{Dierlamm:2012} have been investigated:
 One with $p$-spray and one with $p$-stop implants between the $n^+$\,readout strip implants.
 Fig.\,\ref{fig:Sensor} shows a top view and a cross section of a $p$-stop sensor.
 The layout of the $p$-spray sensors is identical, except that the two narrow $p$-stop implants are replaced by a uniform $p$\,implant, which covers the entire region between the $n^+$-strip implants.

 The sensors were built on 200\,$\upmu$m-thick float-zone (FZ) $\langle 100 \rangle $ silicon with a boron doping of $3.7\times 10^{12}$\,cm$^{-3}$, as determined from capacitance voltage measurements, and an oxygen concentration varying between $3\times 10^{16}$ and $10^{17}\,$cm$^{-3}$, as well as on magnetic-Czochralski (MCz) silicon with similar bulk doping, but with a one order of magnitude higher oxygen concentration\,\cite{Erfle:2014}.
 The 64 $AC$-coupled readout strips each have a length of 25\,mm, a pitch of 80\,$\upmu $m, and are made up of 19\,$\upmu $m-wide $n^+$ implants isolated from the Al strips by 250\,nm of SiO$_2$ and 50\,nm of Si$_3$N$_4$.
 The Al strips overlap the 650\,nm thick SiO$_2$ layers, which cover the region between the strips, by 6\,$\upmu$m, and the entire sensor, with the exception of the bond pads, is covered by an additional 500\,nm of SiO$_2$ for passivation.
 The values for the actual $p$\,dopant concentrations are only poorly known.
 To the best of our knowledge the integrals of the dopant concentrations are $2\times 10^{11}\,$cm$^{-2}$ for the $p$-stop and $5\times 10^{10}\,$cm$^{-2}$ for the $p$-spray implants, respectively.
 The $p$ stops are two 4\,$\upmu$m wide $p^+$ implants at 6\,$\upmu$m separation centered between the readout strips.

\begin{figure}[!ht]
   \centering
   \begin{subfigure}[a]{7.5cm}
    \includegraphics[width=7.5cm]{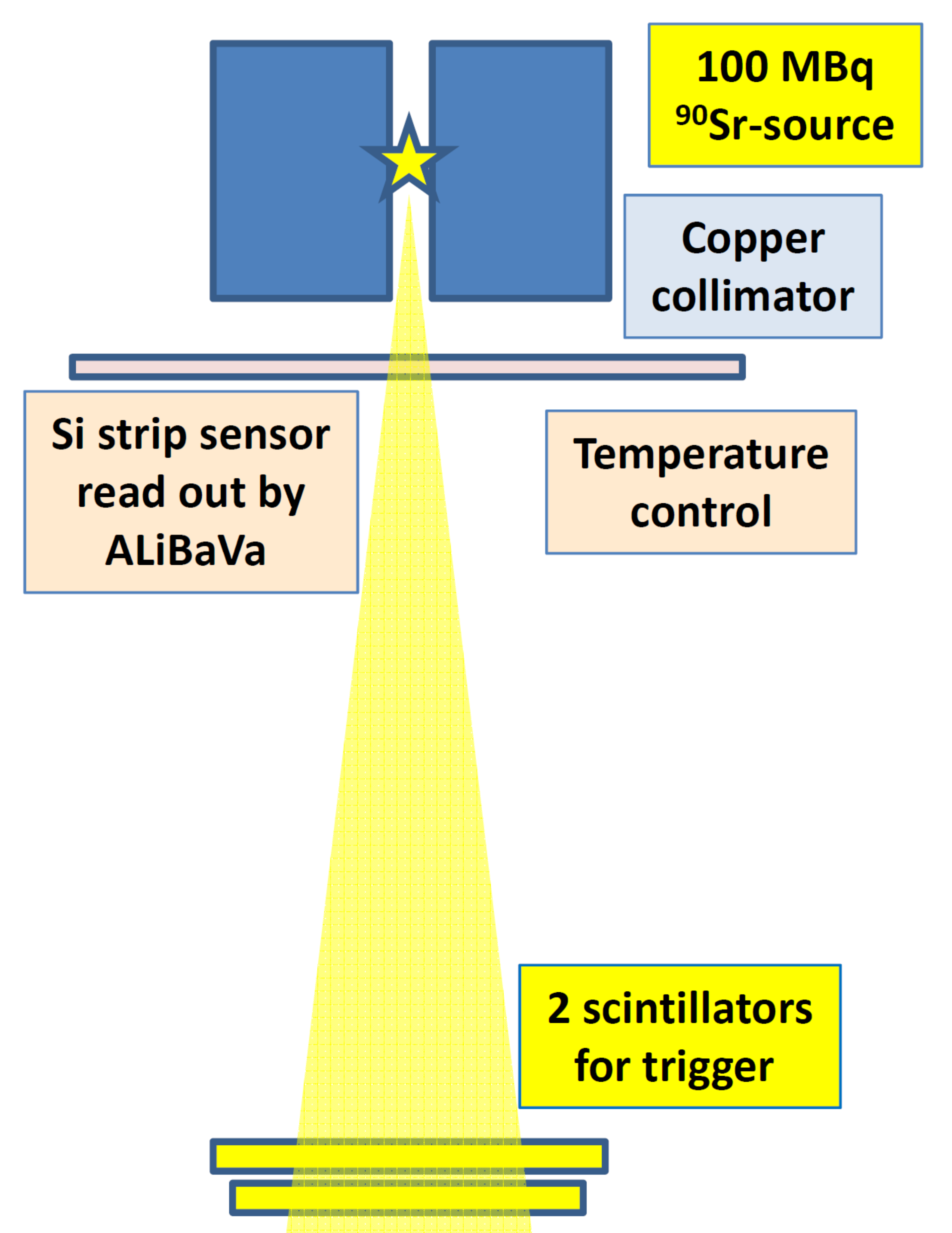}
    \caption{ }
     \label{fig:Setup1}
   \end{subfigure}%
    ~
   \begin{subfigure}[a]{6.8cm}
    \includegraphics[width=6.8cm]{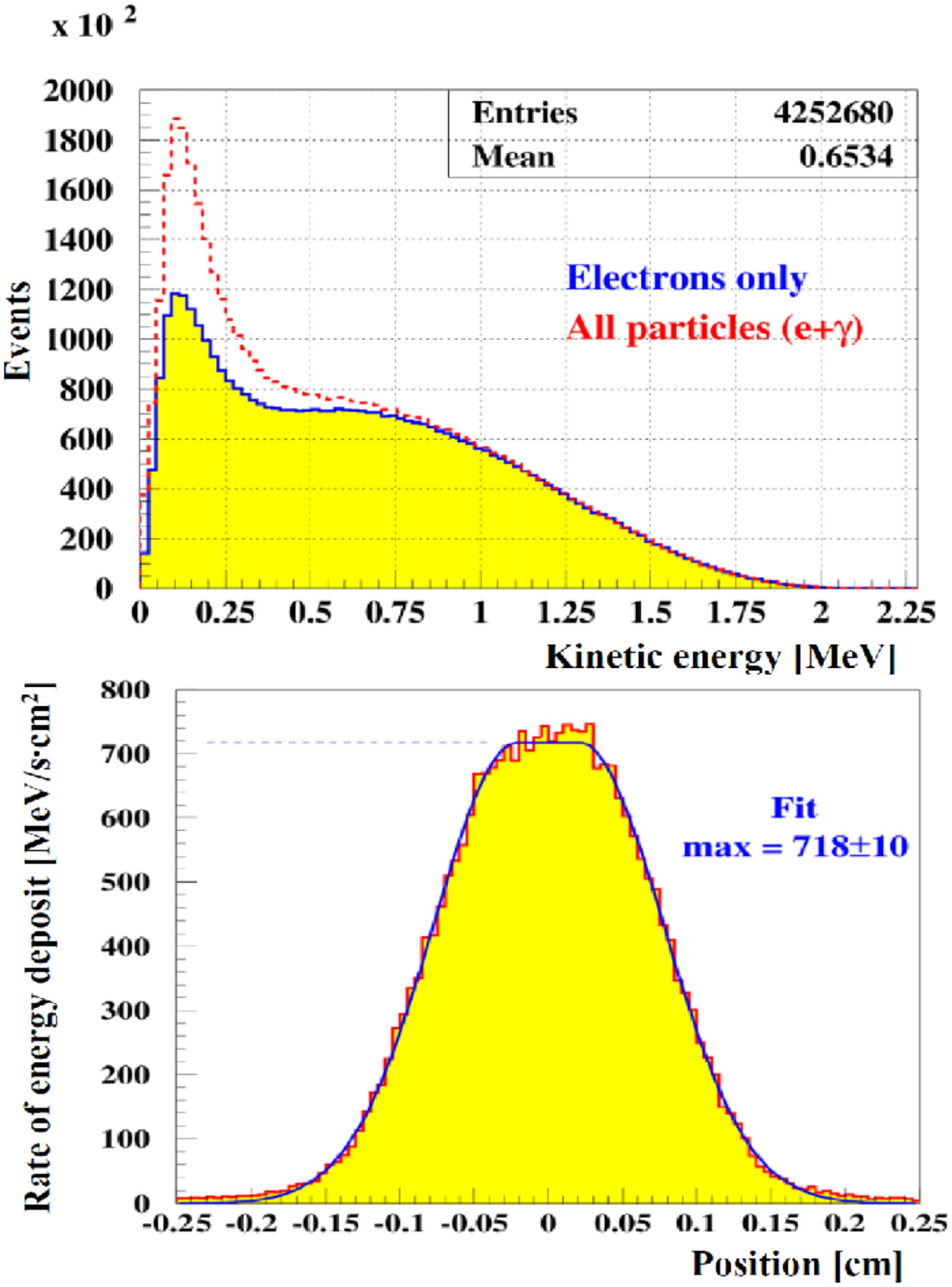}    
    \caption{ }
     \label{fig:Setup2}
   \end{subfigure}%
   \caption{\,(a) Experimental layout. This includes the 100\,MBq $^{90}$Sr\,source placed in a copper collimator, the silicon strip sensor and the two scintillation counters used for triggering. The source can be moved under computer control.
   (b)\,Top: Simulated energy spectrum of the photons and electrons from the $^{90}$Sr\,source at the top of the silicon sensor.
   Bottom: Simulated spatial distribution of the rate of energy-loss density in the SiO$_2$\,layer on top of the sensor, where a thickness of 700\,nm has been assumed in the simulation.
   The distribution of the dose rate is radially symmetric, and a slice through the centre of the distribution is shown.}
  \label{fig:Setup}
\end{figure}

 Fig.\,\ref{fig:Setup1} shows the measurement setup.
 A 100\,MBq $^{90}$Sr\,source, placed in a copper collimator and mounted on a computer controlled $x$-$y$-translation table, irradiated the silicon sensor, which was mounted on a Peltier element for temperature control between $-30\,^\circ $C and $+60\,^\circ $C.
 The sensor was read out by ALiBaVa\,\cite{Alibava:2009}, a multi-channel readout system for silicon strip sensors with a sampling time of 25\,ns. Two 3\,mm-thick plastic scintillators, placed 58 and 62\,mm from the source, respectively, provided the trigger signal from electrons that have traversed the silicon sensor and deposited energy in both scintillators.
 The energy distribution of the electrons and photons and the spatial distribution of the rate of energy-loss density in the SiO$_2$ layer on the surface of the sensor were estimated with a Monte Carlo simulation based on GEANT3 and are shown in Fig.\,\ref{fig:Setup2}.
 The dose-rate distribution is circular with a diameter at full-width-half-maximum of 2\,mm and a value at the maximum of 50\,Gy/d.
 The non-ionizing energy-loss (NIEL) rate, relevant for radiation damage in the silicon bulk, corresponds to about $10^8$\,n$_{eq}$\,/(cm$^{2}\cdot $d).
 The simulated energy-loss distribution of the trigger electrons in the silicon sensor is similar to a Landau distribution for minimum-ionizing-particles (mip), with a most-probable-value (mpv) of 56\,keV, compared to 54\,keV for mips.
 The angular spread of the trigger electrons is about $\pm 100$\,mrad.

 In this paper we study the change of the charge collection as a function of the dose from the $^{90}$Sr\,source of a non-irradiated $p$-stop sensor, a non-irradiated $p$-spray sensor, and a $p$-stop sensor irradiated by $1.5\times 10^{15}$\,n$_{eq}$/cm$^{2}$ 24\,GeV/c protons and $6\times 10^{14}$\,n$_{eq}$/cm$^{2}$ reactor neutrons, which corresponds to an ionizing dose in SiO$_2$ of about 0.75\,MGy.
 This is approximately the radiation exposure expected for the CMS\,experiment at the HL-LHC 15\,cm from the beams after an integrated luminosity of 3\,000\,fb$^{-1}$ has been delivered\,\cite{Mueller:2010, Dawson:2012}.
 Both dose and NIEL rates from the $^{90}$Sr\,source are more than an order of magnitude lower than the maximum rates expected at the HL-LHC.
 If not stated otherwise, the measurements presented in this paper were performed at $-20\,^\circ $C, the temperature at which the silicon sensors are intended to be operated at the HL-LHC.

  \section{Analysis}
    \label{sect:Analysis}

\begin{figure}[!ht]
   \centering
   \begin{subfigure}[a]{7.4cm}
    \includegraphics[width=7cm]{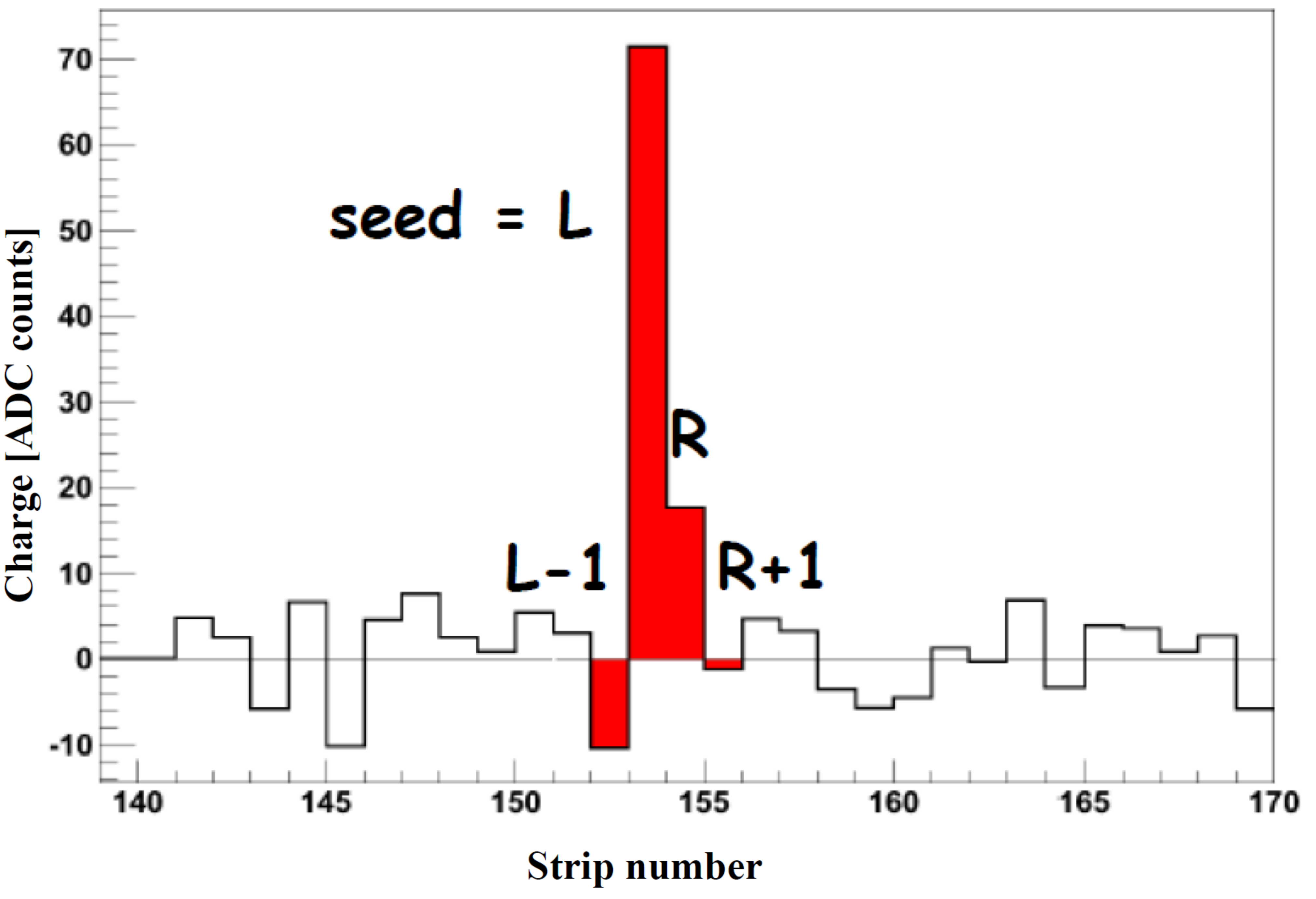}
    \caption{ }
     \label{fig:PH1}
   \end{subfigure}%
    ~
   \begin{subfigure}[a]{7.5cm}
    \includegraphics[width=7.6cm]{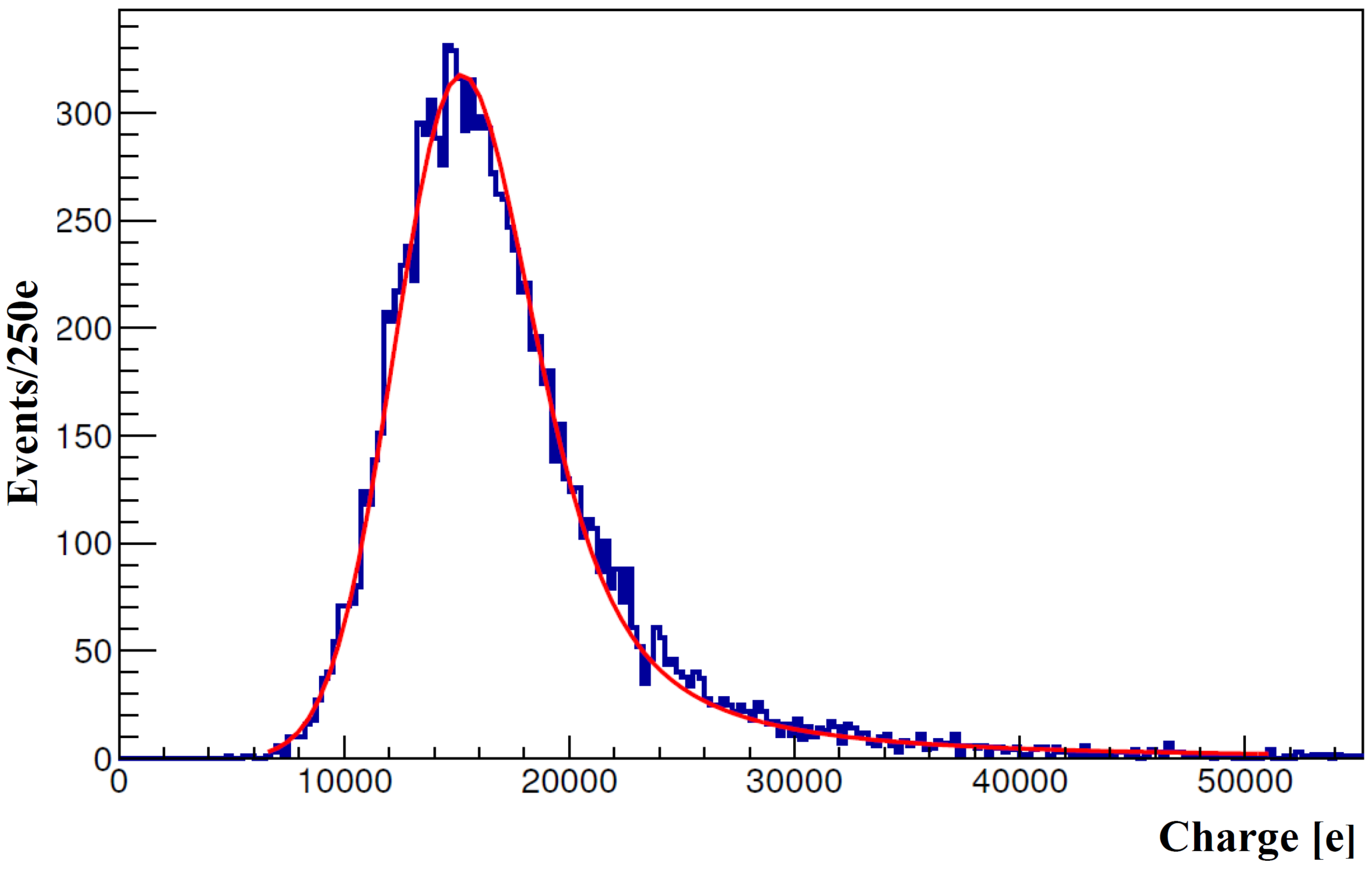}
    \caption{ }
     \label{fig:PH2}
   \end{subfigure}%
   \caption{\,(a) Pulse height versus strip number for a typical event, including an illustration of the naming convention of the strips.
   (b)\,Distribution of the pulse-height sum for the 4 strips of a cluster for the non-hadron-irradiated sensor. The continuous curve represents the fit to the data using a Landau distribution convolved with a Gaussian.}
  \label{fig:PH}
\end{figure}

  In the offline analysis events are selected if the trigger signal was in phase within $\pm 5$\,ns with the 40\,MHz ALiBaVa clock.
  Fig.\,\ref{fig:PH1} shows the pulse height (PH) versus strip number after pedestal and common mode subtraction for a typical event\,\cite{Erfle:2014}.
  The strip with the largest PH is called the seed, whilst its largest neighbour defines the region of passage of the electron. These 2 strips, together with the 2 next-to-next neighbours are the 4 strips used in the analysis of an event.
  They are labeled $L$-1, $L$, $R$, and $R$+1, respectively.
  Fig.\,\ref{fig:PH2} shows the distribution of the sum of the pulse heights of the 4 strips, PH(4-cluster), in units of electron charge, e, for 5\,000 events measured using a non-hadron-irradiated sensor after pedestal and common-mode subtraction.
  The calibration\,\cite{Erfle:2014} uses the charge injection feature of the ALiBaVa readout system, which has an accuracy of better than 1\,\%.
  As expected from the simulation, the distribution can be fitted using the convolution of a Landau distribution with a Gaussian, which is shown as the smooth curve in Fig.\,\ref{fig:PH2}.
  In subsequent analyses, however, we will use the median, as for individual strips the pulse-height distributions cannot be described by the convolution of Landau and Gaussian distributions.
  The noise, with a root-mean-square (rms) value of about 810\,e, has a good separation from the electron signal, which has an mpv of about  16\,000\,e.
  This mpv value is compatible with the simulated most probable energy loss of 56\,keV.
  After hadron irradiation the rms noise increased to about 950\,e.

 The analysis  uses the variable $\eta = PH(R)/(PH(R)+PH(L))$ introduced in Ref.\,\cite{Belau:1983}.
 The d$N$/d$\eta $\,distribution allows the investigation of the electric field distribution, as well as the charge-sharing and charge-loss properties of segmented sensors.
 The d$N$/d$\eta $\,distribution will be the sum of two $\delta $-functions, one at $\eta = 0$ and the other at $\eta = 1$ under the following conditions:
 All field stream lines originate at the $n^+$\,implants of the readout strips, the readout noise is zero, charge diffusion is neglected, and the particles traverse the sensor at normal incidence.
 Electronic cross-talk shifts the positions of the $\delta $-functions inwards, and noise causes a broadening and a further inward shift.
 Diffusion, which broadens the charge distribution arriving at the readout strips by a few $\upmu$m, results in some charge sharing. An angular spread of the traversing particles further increases charge sharing.
 If some field stream lines originate at the Si-SiO$_2$ interface, as shown in some of the simulations discussed in Sect.\,\ref{sect:Discussion}, charge sharing will increase further, and the d$N$/d$\eta $ distribution in between the peaks at low and high $\eta $ values will be populated.
 The more field lines originate at the Si-SiO$_2$ interface, the more events will appear in the central $\eta $\,region.

\begin{figure}[!ht]
   \centering
    \begin{subfigure}[a]{7.5cm}
    \includegraphics[width=7.5cm]{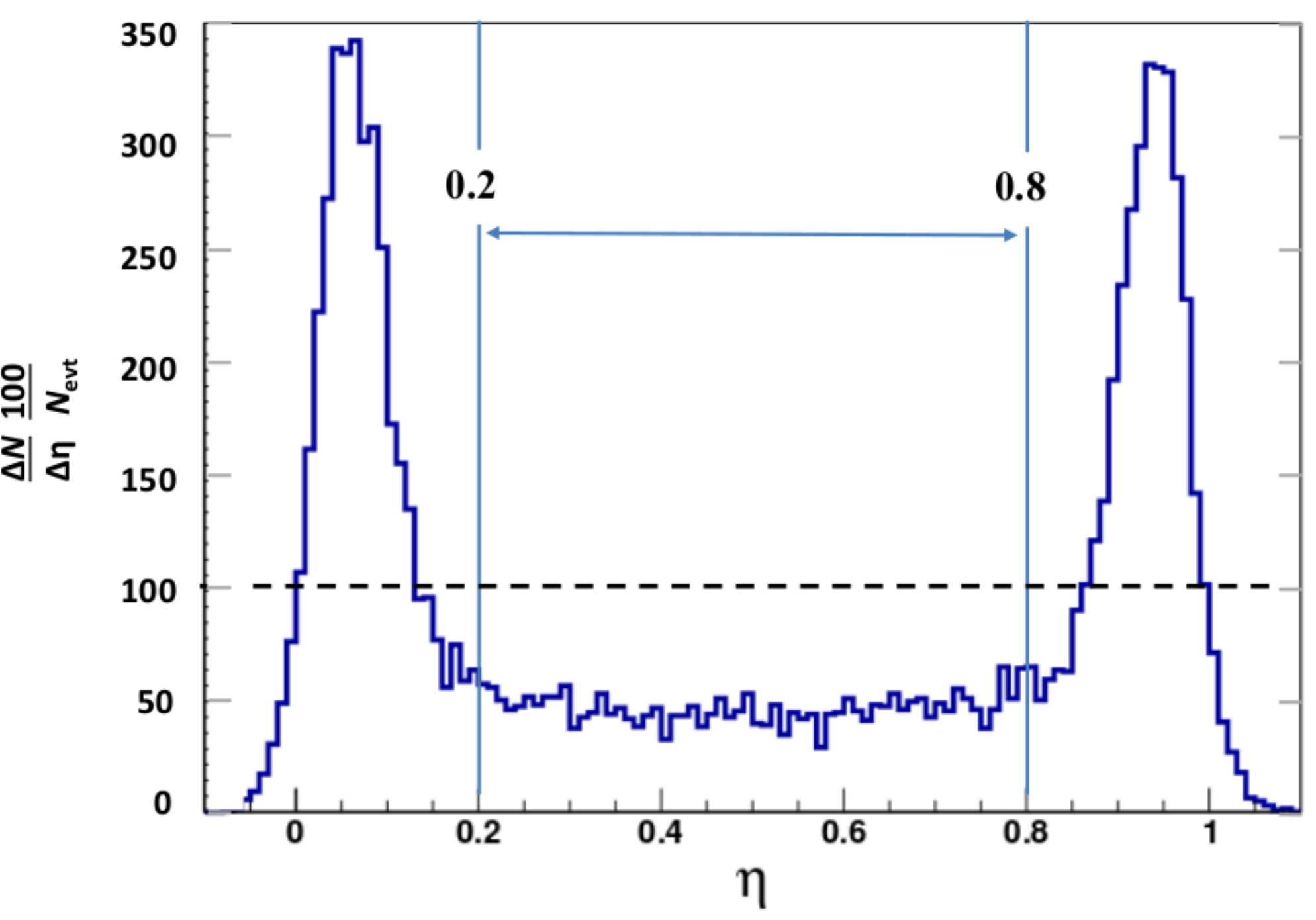}
    \caption{ }
     \label{fig:Eta1}
   \end{subfigure}%
    ~
   \begin{subfigure}[a]{7.5cm}
    \includegraphics[width=7.1cm]{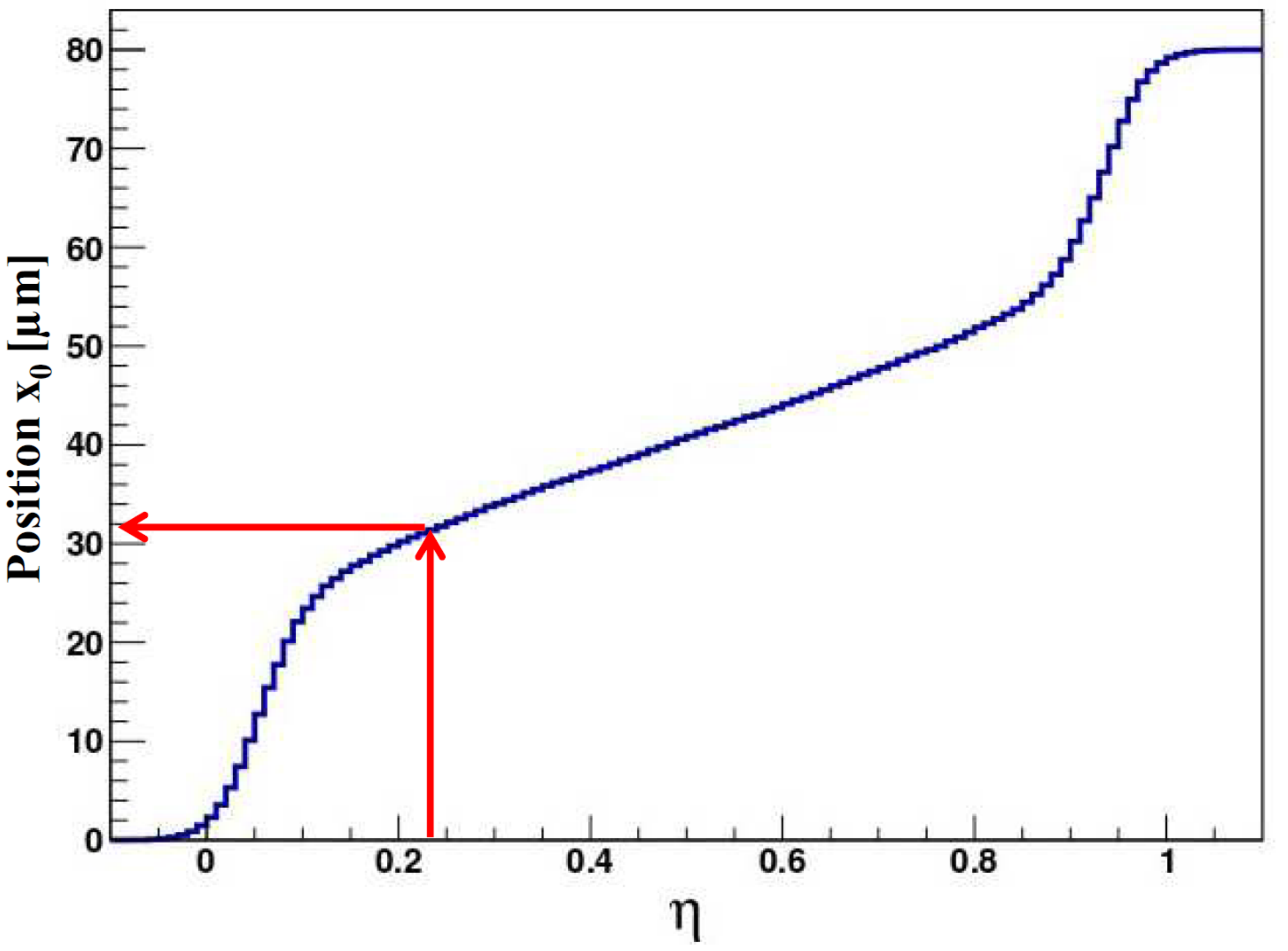}
    \caption{ }
     \label{fig:Eta2}
   \end{subfigure}%
   \caption{\,(a) An example of the differential distribution
   $(\Delta N/\Delta \eta )\times 100/N_{evt}.$
   (b)\,Normalized cumulative $\eta $ distribution, which relates $\eta $ and $x_0$, the distance of the particle passage from the centre of strip $L$.}
  \label{fig:Eta}
\end{figure}

 Fig.\,\ref{fig:Eta1} shows the distribution of
 $(\Delta N/\Delta \eta )\times 100/N_{evt}$
 measured for the non-hadron-irradiated $p$-stop sensor biased at 600\,V after 0.2\,days of $\beta $\,source irradiation, which corresponds to a dose of 10\,Gy.
 $N_{evt}$ is the total number of events and $\Delta N$ the number of events in a bin of width $\Delta \eta$.
 The value found in the central $\eta $ region is $45.2 \pm 1.2 \,\%$, which is significantly higher than the $15 - 20 \,\%$ expected from the angular spread of $\pm 100$\,mrad.
 To characterize charge sharing we use the quantity
 $CS = 100 \times (\Delta N(0.2 - 0.8)/(0.6 \times N_{evt}))$,
 where $\Delta N(0.2 - 0.8)$ is the number of events in the interval $ 0.2 < \eta < 0.8$, whilst 0.6 is the width of the $\eta $\,interval.
 Thus $CS$ gives the percentage of charge sharing relative to 100\,\% charge sharing.

 In addition, $\eta $ allows the determination of $x_0$, the distance of the traversing particle from the centre of strip $L$, as discussed in Ref.\,\cite{Belau:1983}.
 The following is a brief summary of the derivation.
 Assuming a uniform distribution of $N_{evt}$ events over the sensor with inter-strip spacing $pitch$, the fraction of events in a given interval $\Delta \eta $, $(1/N_{evt}) \times (\mathrm{d} N/\mathrm{d} \eta ) \times \Delta \eta $, is equal to $\Delta x_0/pitch $,
 where $\Delta x_0$ is the $x_0$\,interval that corresponds to the selected $\Delta \eta $ interval.
 Thus the cumulative distribution normalized to \textit{pitch} relates the measured $\eta $\,value to the distance $x_0$ of the particle from strip $L$.
 Fig.\,\ref{fig:Eta2} shows an example of a measured $x_0 -\eta $ relation.

 The position resolution $\delta x$ as a function of $x_0$ can be estimated using $\delta x = (\mathrm{d}x_0/\mathrm{d} \eta ) \times \delta \eta$, where the uncertainty $\delta \eta$ can be calculated from the signal-to-noise-ratio and the definition of $\eta$.
 For a flat $\mathrm{d} N/\mathrm{d} \eta$ distribution between $\eta = 0$ and 1 the $x_0 - \eta $ relation is linear, and $\delta x$ is independent of $x_0$.
 We call this \textit{ideal charge division}, if in addition most of the signal is induced in strips $L$ and $R$.

  \section{Results}
    \label{sect:Results}

 We first present results for the two non-hadron-irradiated sensors and then for the hadron-irradiated $p$-stop sensor. An explanation and discussion of the observations with the aid of SYNOPSYS TCAD\,\cite{SYNOPSYS} simulations that include surface charges at the Si-SiO$_2$ interface and different boundary conditions on the sensor surface are given in Sect.\,\ref{sect:Discussion}.

 \begin{figure}[!ht]
   \centering
   \begin{subfigure}[a]{7.5cm}
    \includegraphics[width=7.5cm]{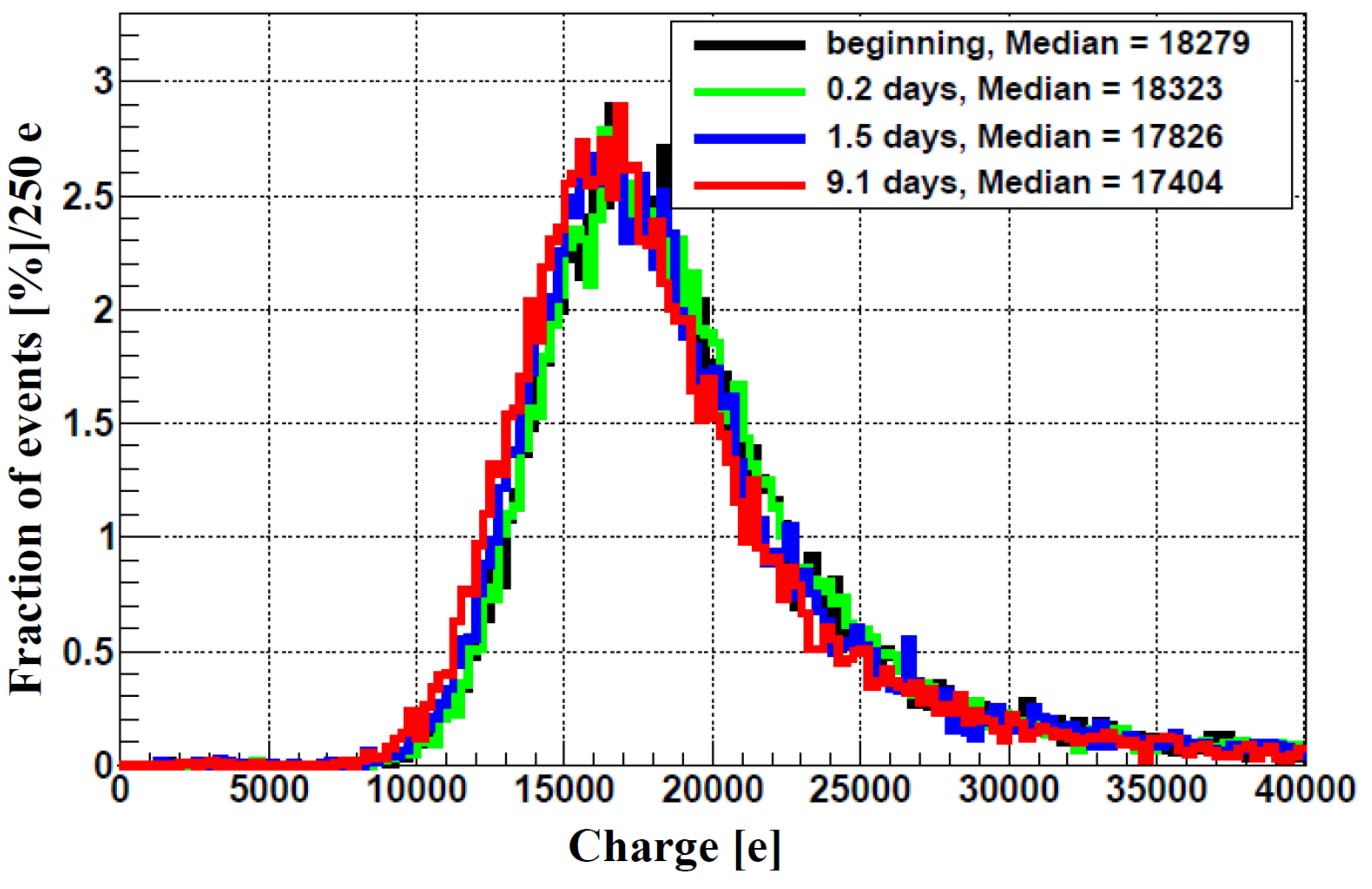}
    \caption{ }
     \label{fig:Ph1}
   \end{subfigure}%
    ~
   \begin{subfigure}[a]{7.5cm}
    \includegraphics[width=7.5cm]{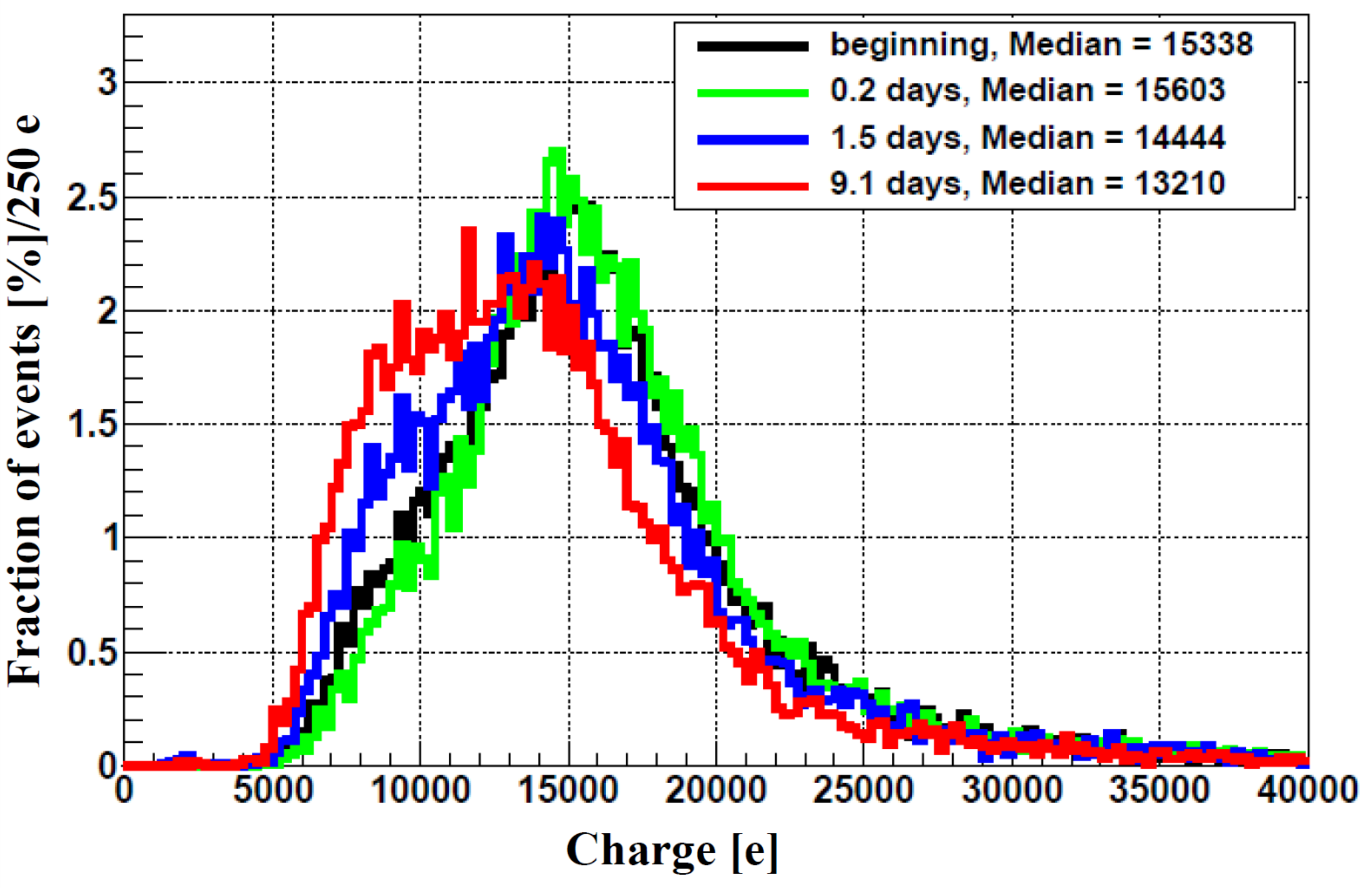}
    \caption{ }
     \label{fig:Ph2}
   \end{subfigure}%
   \caption{\,(a) PH(4-cluster) distributions for the non-hadron-irradiated MCz $p$-stop sensor biased at 600\,V after exposure to $\beta $\,doses of 0\,Gy (start of measurements), 10\,Gy (after 0.2\,days), 75\,Gy (after 1.5\,days), and 450\,Gy (after 9.1\,days).
   (b)\,The corresponding PH(seed) distributions.}
  \label{fig:Ph}
\end{figure}

 Fig.\,\ref{fig:Ph1} shows the pulse-height distributions of the 4-strip clusters, PH(4-cluster), whilst Fig.\,\ref{fig:Ph2} shows the corresponding pulse-height distributions for the seed clusters, PH(seed), for the non-hadron-irradiated $p$-stop sensor fabricated on magnetic-Czochralski material, as measured 0, 0.2, 1.5 and  9.1 days after the start of irradiation with the $\beta $\,source.
 The corresponding doses in the SiO$_2$ are 0, 10, 75 and 450\,Gy, respectively.
 For PH(4-cluster) we observe Landau distributions with dose-dependent changes of the median by $ +0.25, -2.5$ and $-4.8$\,\%, respectively.
 Significantly larger changes are observed for PH(seed):
 The median changes by $ +1.7, -5.8$ and $-14.0$\,\%, respectively.
 In addition, the shape of the PH(seed) distribution changes:
 The approximately triangular distribution with a maximum around 15\,000\,e changes to a distribution that is nearly flat between 7\,500 and 15\,000\,e.

 \begin{figure}[!ht]
   \centering
   \begin{subfigure}[a]{7.5cm}
    \includegraphics[width=7.5cm]{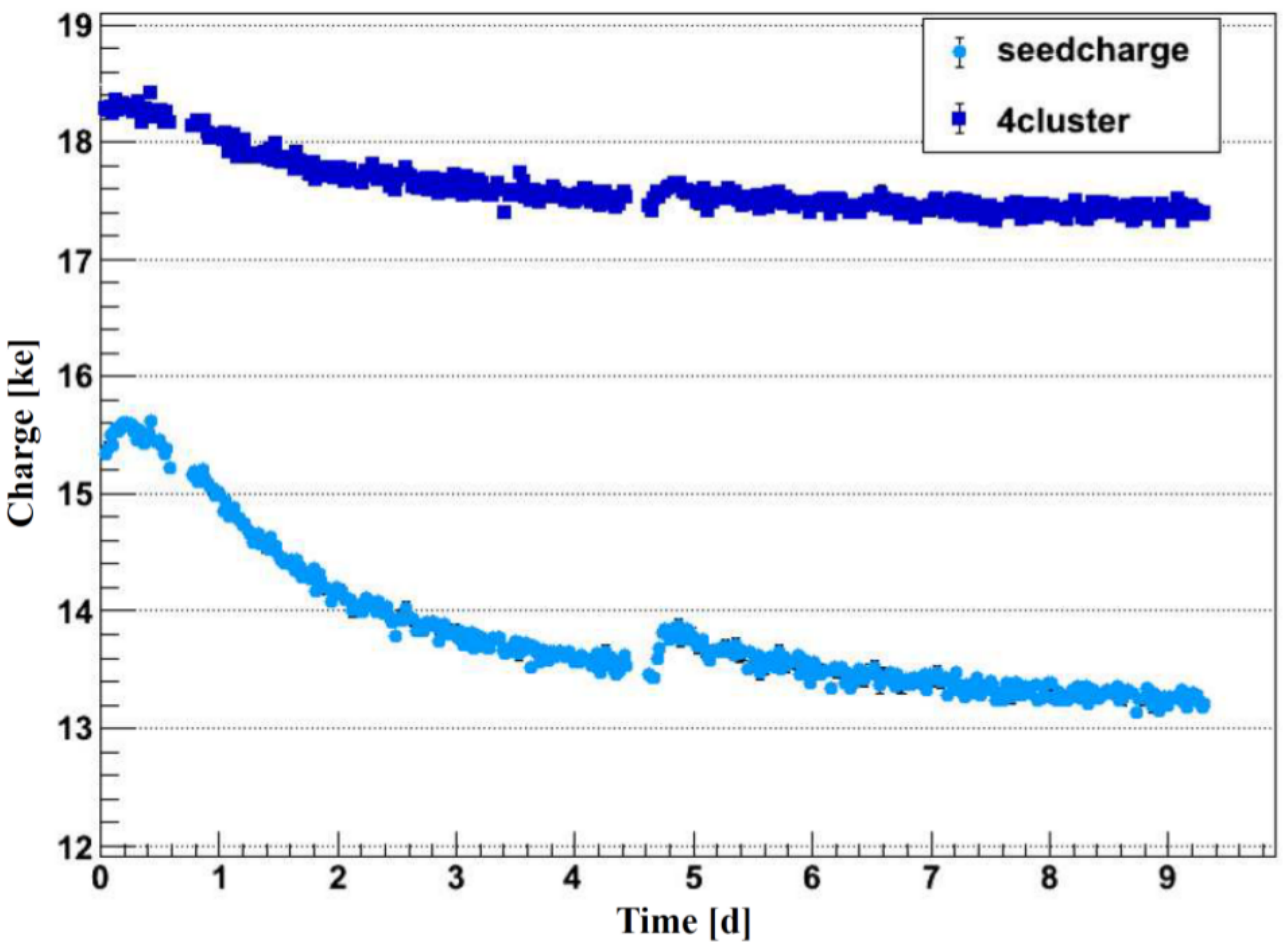}
    \caption{ }
     \label{fig:Pstop1}
   \end{subfigure}%
    ~
   \begin{subfigure}[a]{7.5cm}
    \includegraphics[width=7.5cm]{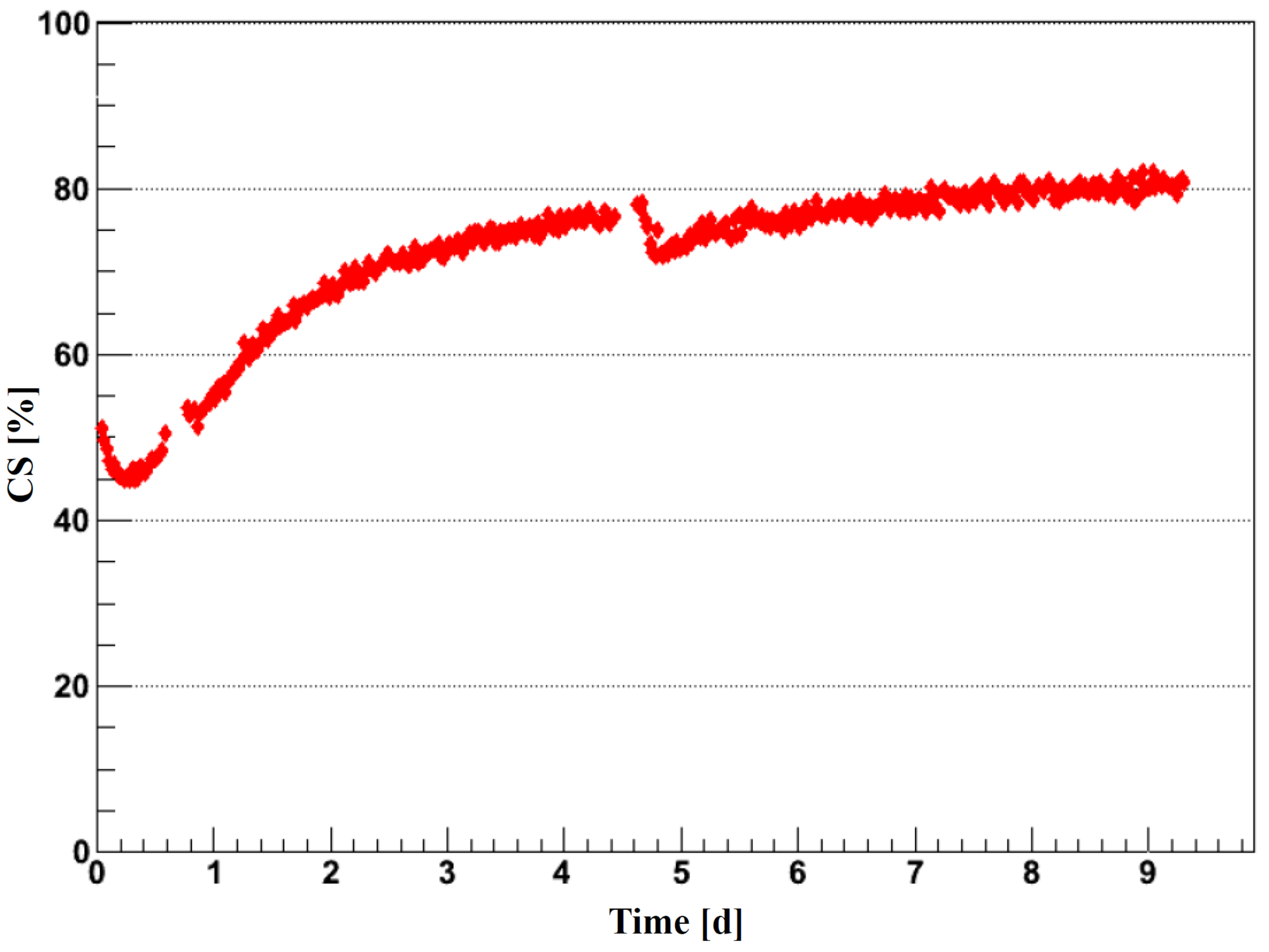}
    \caption{ }
     \label{fig:Pstop2}
     \end{subfigure}%
    \caption{\,(a) Median of the PH(4-cluster) and PH(seed) distributions for the non-hadron-irradiated MCz $p$-stop sensor biased at 600\,V as a function of the measurement time.
    The initial dose is 0\,Gy and the dose after 9\,days  about 450\,Gy.
    At 4.6 days calibration runs were taken, and the sensor was not exposed to the $\beta $\,source.
   (b)\,The corresponding time dependence for $CS$.}
  \label{fig:Pstop}
\end{figure}

 Fig.\,\ref{fig:Pstop1} shows the time dependence of the median of the PH(4-cluster) and PH(seed) distributions for the non-hadron-irradiated MCz $p$-stop sensor biased at 600\,V, whilst Fig.\,\ref{fig:Pstop2} shows the corresponding dependence of the charge sharing, $CS$.
 As a function of the measurement time, the dose in the SiO$_2$ increases from 0 to about 450\,Gy.
 For PH(4-cluster) we observe a constant value up to about 0.4\,days, and then a steady decrease by about 5\,\%.
 For PH(seed) a much larger decrease by 15\,\% after an initial short term increase of 3\,\% is observed.
 $CS$ is initially 52\,\%, 45\,\% after 0.3\,days (15\,Gy), and then steadily increases to 80\,\%.
 After 4.6\,days the $\beta $\,source was retracted, and a calibration of the electronics and a voltage scan for pedestal and noise determination between 0 and 1000\,V was performed.
 This is seen as gaps and steps in the time dependencies.

 \begin{figure}[!ht]
  \centering
   \includegraphics[width=7.5cm]{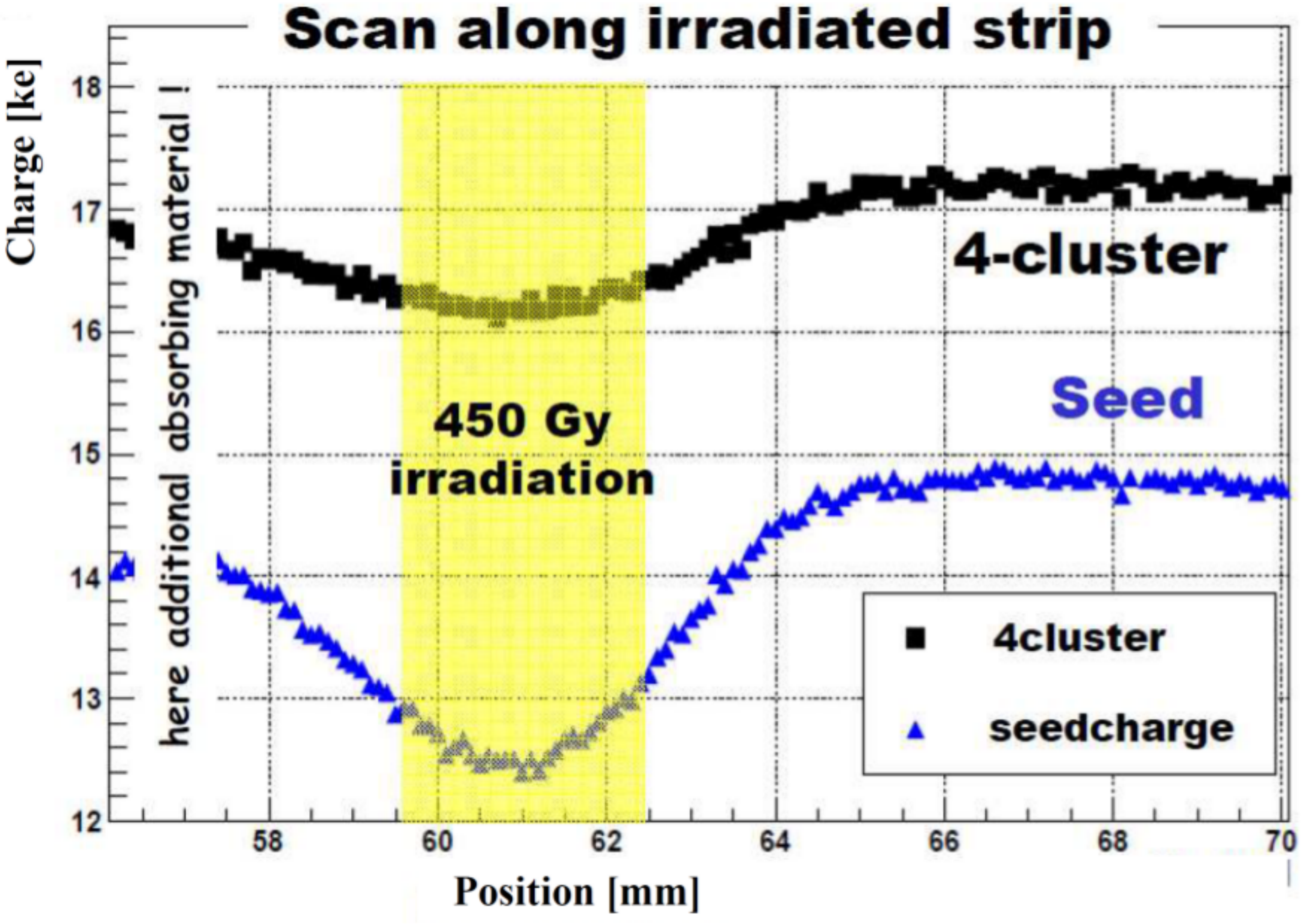}
    \caption{Source scan along the strips on which the $\beta $\,source had been centred and the sensor irradiated to a dose of 450\,Gy. The yellow band indicates the region of the highest dose.}
   \label{fig:StripScan}
 \end{figure}

 A source scan along the strips on which the source had been centred for 9.5\,days is shown in Fig.\,\ref{fig:StripScan}.
 It is found that the decrease in PH(4-cluster) and in PH(seed) is limited to the region where the source had been positioned and that outside this region no effects of the irradiation are observed.

 To investigate a possible dose-rate dependence, measurements were performed with a 38\,MBq $^{90}$Sr\,source at a position that had not previously been exposed to the $\beta $\,source.
 It is found that, within the accuracy of the measurements, the results only depend on dose and not on the dose-rate\,\cite{Erfle:2014}.

 \begin{figure}[!ht]
   \centering
   \begin{subfigure}[a]{7.5cm}
    \includegraphics[width=7.5cm]{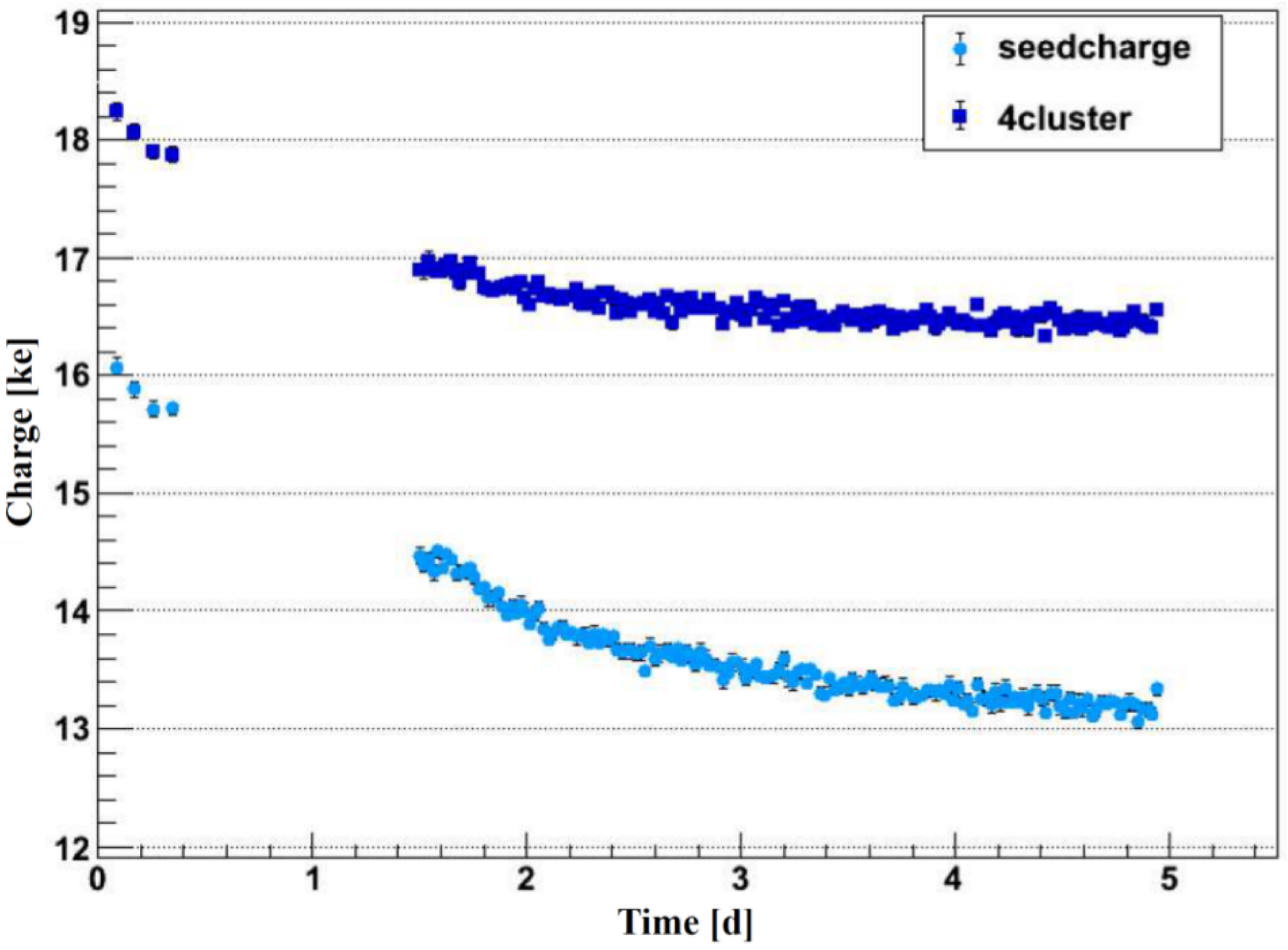}
    \caption{ }
     \label{fig:Pspray1}
   \end{subfigure}%
    ~
   \begin{subfigure}[a]{7.5cm}
    \includegraphics[width=7.5cm]{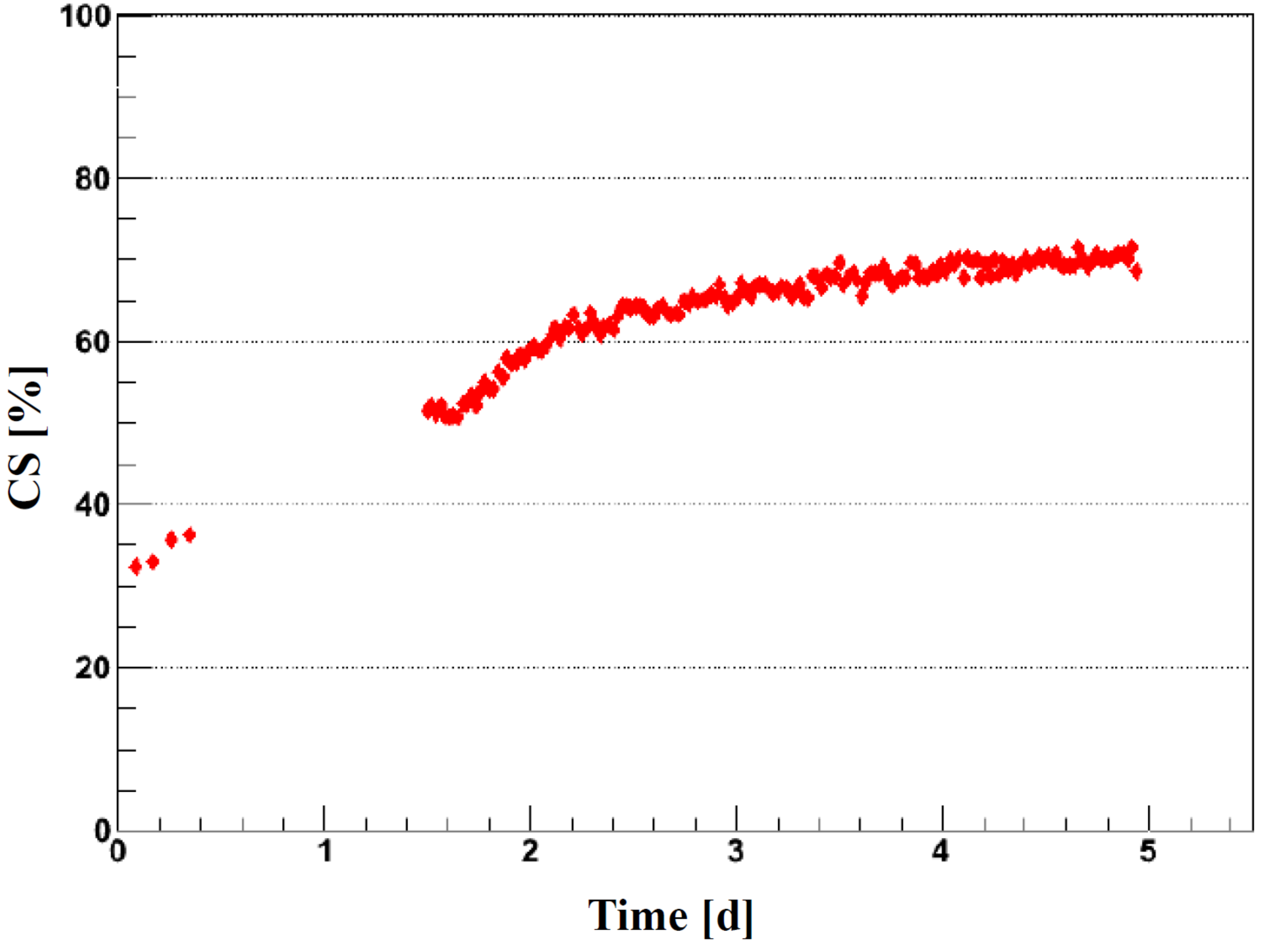}
    \caption{ }
     \label{fig:Pspray2}
   \end{subfigure}%
   \caption{\,(a) Median of the PH(4-cluster) and PH(seed) distributions for the non-hadron-irradiated $p$-spray sensor biased at 600\,V  as a function of the time from the beginning of the irradiation. The initial dose is 0\,Gy and the dose after 5\,days,  250\,Gy.
   Between 0.4 and 1.4\,days the measurement was interrupted because of a problem with the nitrogen flow, which is needed to prevent ice formation on the sensor at $-20\,^\circ $C. During that time the sensor remained exposed to the $\beta $\,source.
   (b)\,The corresponding time dependence for $CS$.}
  \label{fig:Pspray}
\end{figure}

 Fig.\,\ref{fig:Pspray} shows the time dependence of the median values of PH(4-cluster) and PH(seed), and of $CS$ measured for the non-hadron-irradiated $p$-spray sensor built on float-zone silicon.
 Data between 0.4 and 1.4\,days are missing, as during this time the nitrogen flow was interrupted and the sensor covered by an ice layer.
 Qualitatively the results are similar to those from the non-hadron-irradiated $p$-stop sensor:
 The median of PH(4-cluster) decreases by about 10\,\%, PH(seed) by 20\,\%, and $CS$ increases from 30\,\% to 70\,\%.
 We note the absence of the initial short time change observed for the $p$-stop sensor.

 \begin{figure}[!ht]
   \centering
   \begin{subfigure}[a]{7.5cm}
    \includegraphics[width=7.5cm]{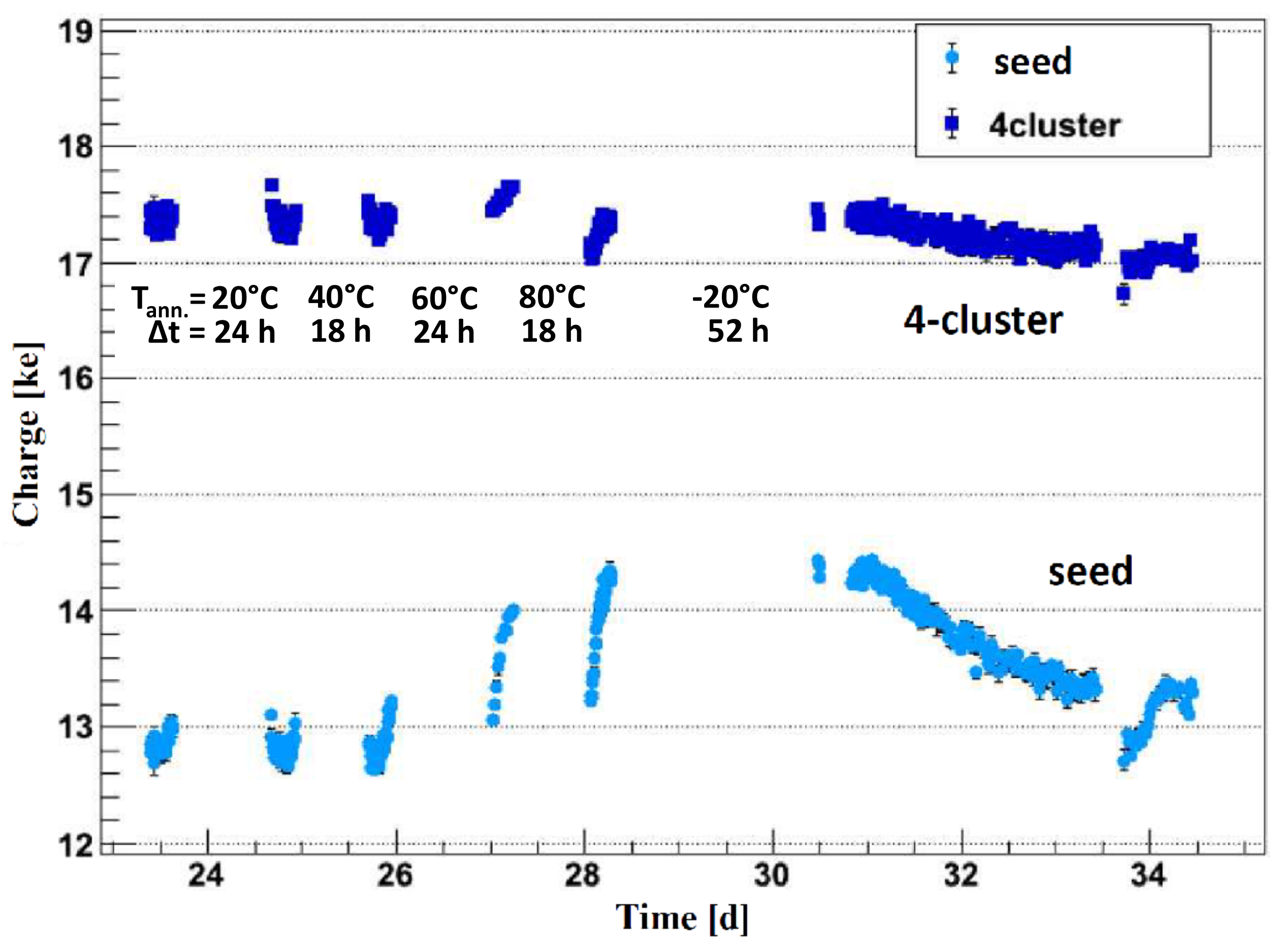}
    \caption{ }
     \label{fig:Anneal1}
   \end{subfigure}%
    ~
   \begin{subfigure}[a]{7.5cm}
    \includegraphics[width=7.5cm]{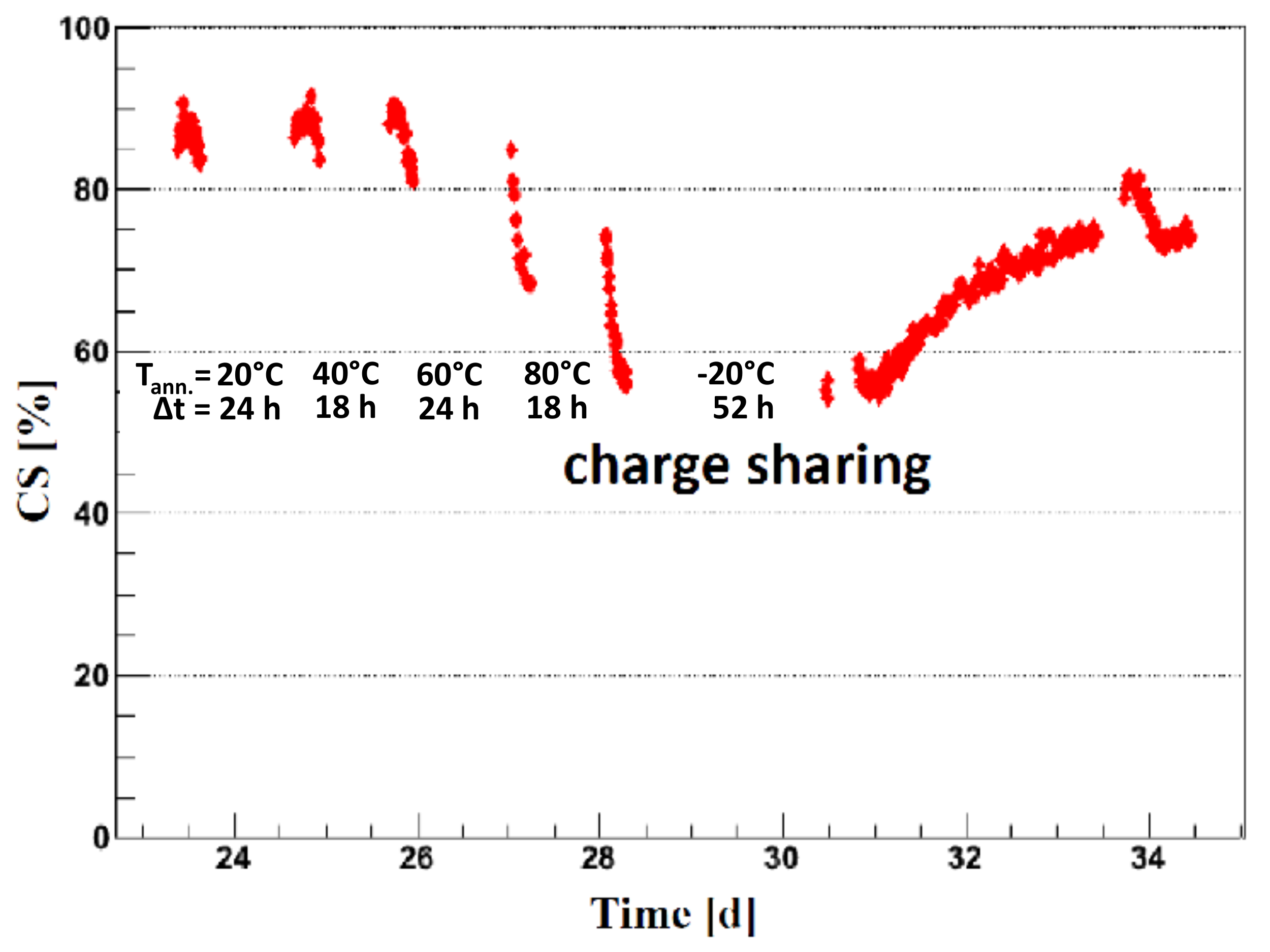}
    \caption{ }
     \label{fig:Anneal2}
   \end{subfigure}%
   \caption{\,(a) Median of the PH(4-cluster) and PH(seed) distributions for the MCz $p$-stop sensor biased at 600\,V after irradiation by the $\beta $\,source to a dose of 800\,Gy as a function of time for the annealing steps (annealing time $\Delta t$, annealing temperature $T_{ann.})$ shown in Table\,\ref{tab:annealing} and discussed in the text.
   (b)\,The corresponding time dependence for $CS$.}
  \label{fig:Anneal}
\end{figure}

 \begin {table}[!ht]
  \centering
   \caption{Annealing history of the non-hadron-irradiated MCz $p$-stop sensor. The \textit{End time of annealing} refers to the time since the beginning of the irradiation used for the $x$\,axes of the distributions in Fig.\,\ref{fig:Anneal}.}
    \vspace{2mm}
 \begin{tabular}{ c c c c c c c}
   \hline
  Annealing step & 1 & 2 & 3 & 4 & 5 & 6 \\
     \hline
  End time of annealing [d] & 23.4 & 24.7 & 25.7 & 27.0 & 28.1 & 30.5  \\
  Duration of annealing [h] & 120 & 24 & 18 & 24 & 18 & 52 \\
  SiO$_2$ dose [Gy] before annealing & 780 & 790 & 810 & 820 & 830 & 850 \\
  Temperature [$^\circ $C] & $-20$ & 20 & 40 & 60 & 80 & $-20$ \\
  Bias voltage [V] during annealing & 600 &  600 & 600 &  600 & 350&  600 \\
   \hline
   \end{tabular}
  \label{tab:annealing}
 \end{table}

 \begin{figure}[!ht]
   \centering
   \begin{subfigure}[a]{7.5cm}
    \includegraphics[width=7.5cm]{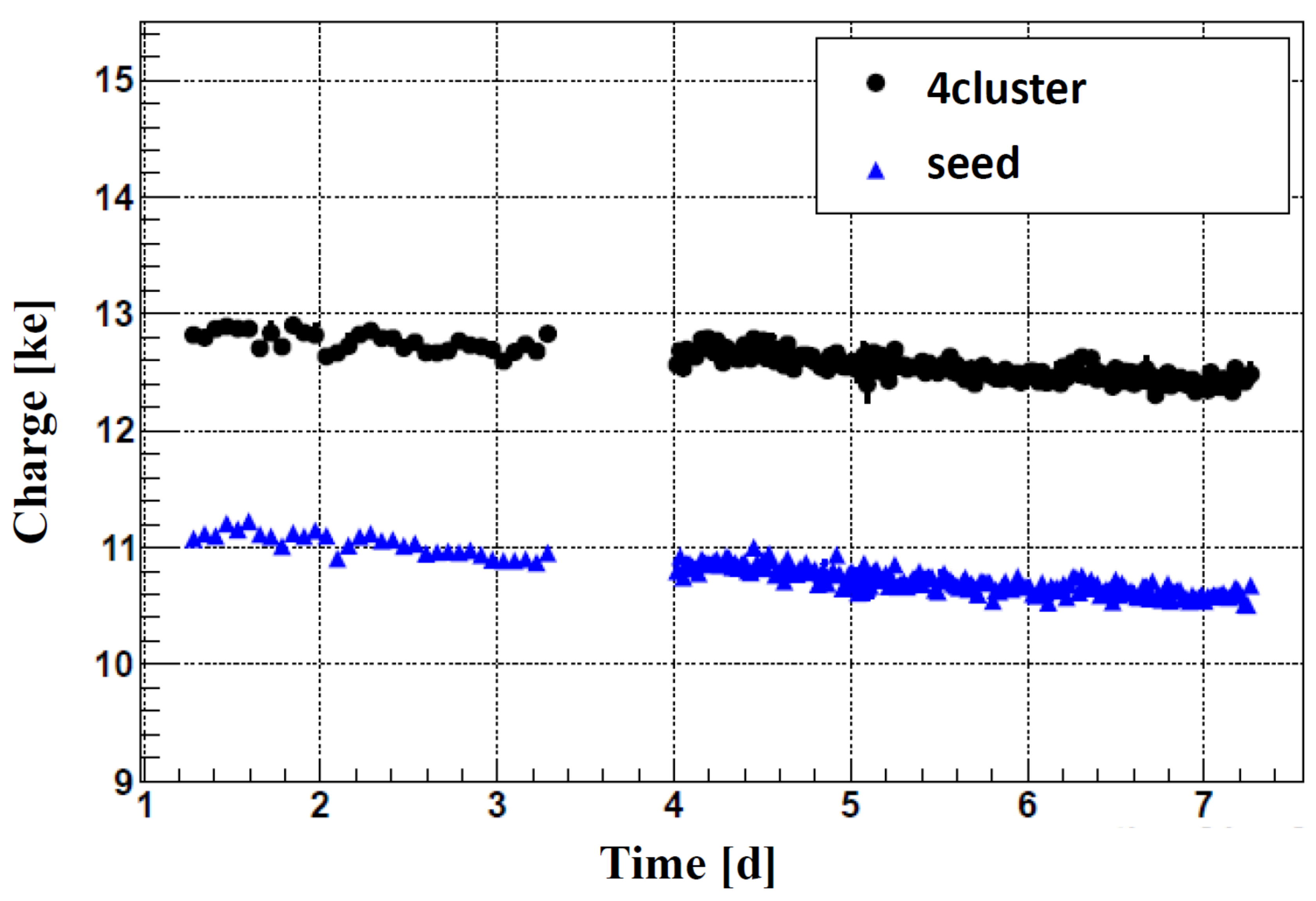}
    \caption{ }
     \label{fig:Irrad1}
   \end{subfigure}%
    ~
   \begin{subfigure}[a]{7.5cm}
    \includegraphics[width=7.5cm]{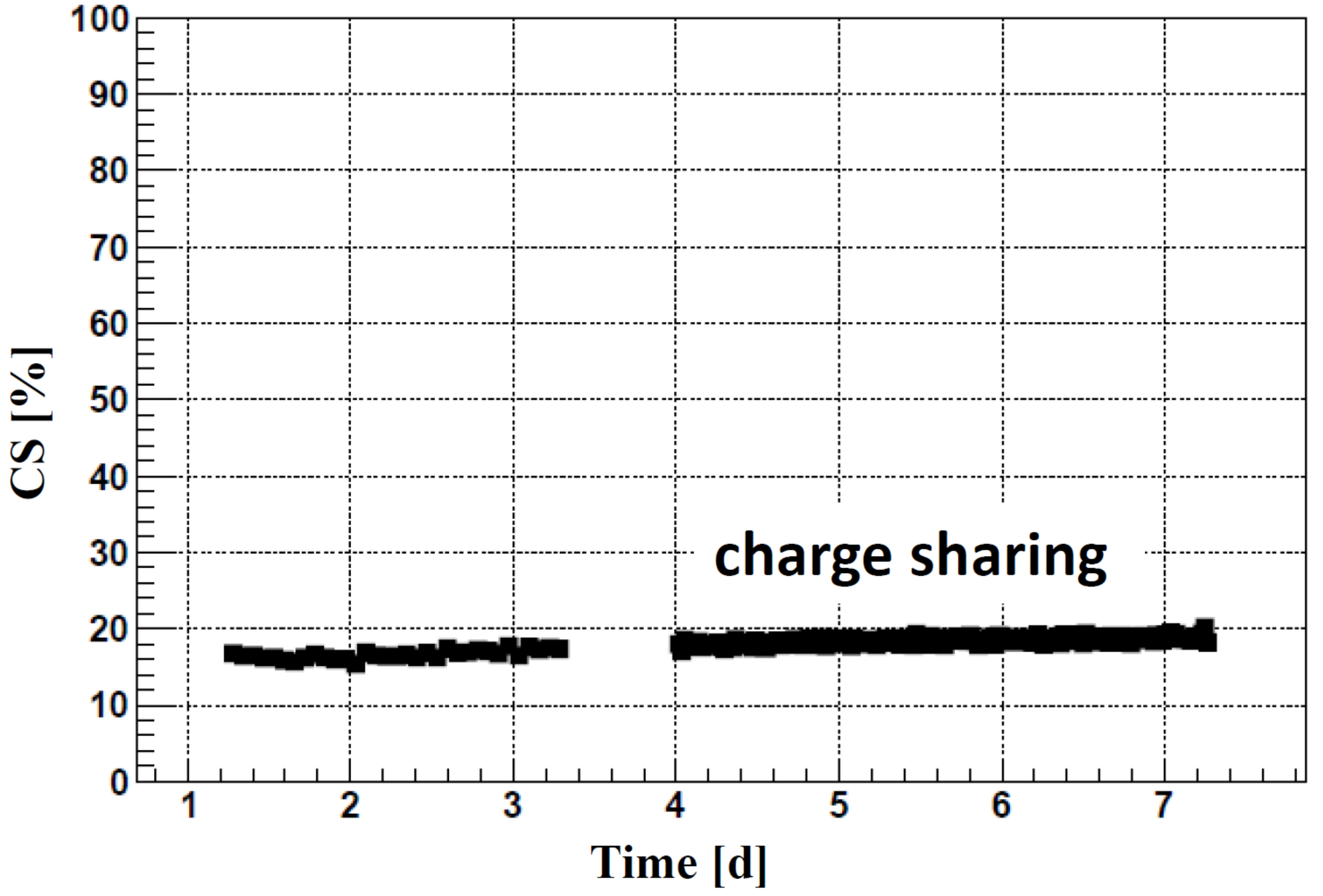}
    \caption{ }
     \label{fig:Irrad2}
   \end{subfigure}%
    \caption{\,(a) Median of the PH(4-cluster) and PH(seed) distributions for the FZ $p$-stop sensor irradiated by $15\times 10^{14}$\,n$_{eq}$/cm$^2$ 24\,GeV/c protons and $6\times 10^{14}$\,n$_{eq}$/cm$^2$ reactor neutrons as a function of the measurement time, during which the dose in the SiO$_2$ from the $\beta $\,source increased from 0 to 375\,Gy.
    The applied voltage was 1000\,V.
   (b)\,The corresponding time dependence for $CS$.}

  \label{fig:Irrad}
\end{figure}

  Next, the annealing behaviour was  investigated for the $p$-stop sensor fabricated using magnetic-Czochralski silicon after a dose of about 800\,Gy from the $\beta $\,source.
 Table\,\ref{tab:annealing} lists the individual annealing steps, and Fig.\,\ref{fig:Anneal} shows the measured median values of PH(4-cluster), PH(seed) and $CS$.
 Only small changes are observed for PH(4-cluster):
 The largest change is a decrease by about 3\,\% after annealing at 80\,$^\circ $C, which quickly recovers after a few hours of exposure to the $\beta $\,source.
 The changes for PH(seed) are larger:
 After every annealing step, PH(seed) is seen to have reached a value close to the one before annealing.
 During the typical 6\,h of irradiation with the $\beta $\,source immediately after the annealing, PH(seed) increases by $\approx 8\,\%$ for the annealing for 24\,h at 60\,$^\circ $C, and $\approx 11\,\%$ for 18\,h at 80\,$^\circ $C.
 Finally, during the 3\,days of long-term measurements, PH(seed) decreases again and approaches the steady-state values shown in Fig.\,\ref{fig:Pstop} at 9 days.
 The charge sharing $CS$ also shows significant changes:
 After annealing, the value of $CS$ has increased, however it decreases quickly when the sensor is irradiated again.
 After a long-term exposure of 3\,days, the steady-state values shown in Fig.\,\ref{fig:Pstop} at 9 days, are approached.
 It is clear that the observed behaviour is quite complex.

 Finally, Fig.\,\ref{fig:Irrad} gives results for the hadron-irradiated $p$-stop sensor fabricated on float-zone silicon measured at $-20\,^\circ $C and a voltage of 1000\,V:
 The median of PH(4-cluster) decreases by about 3\,\% after 7.5\,days of irradiation, corresponding to a $\beta $\,dose in the SiO$_2$ of 375\,Gy.
 The median of PH(seed) decreases by about 5\,\% and $CS$ increases from 16\,\% to 19\,\%.
 We note that for the hadron-irradiated sensor the effects are significantly smaller than for the non-hadron-irradiated sensors.

  \begin{figure}[!ht]
   \centering
   \begin{subfigure}[a]{7.5cm}
    \includegraphics[width=7.5cm]{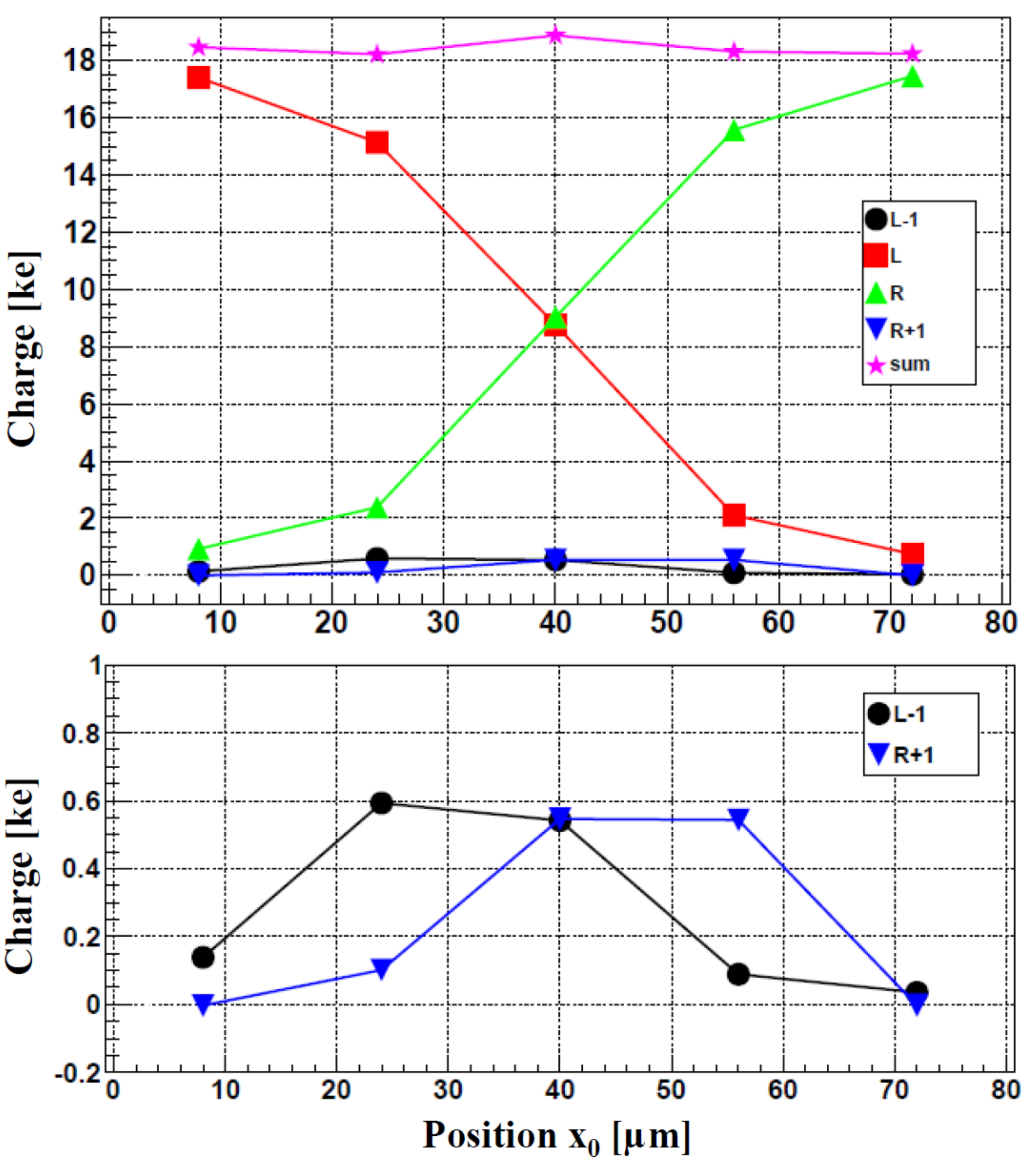}
    \caption{ }
     \label{fig:PHx1}
   \end{subfigure}%
    ~
   \begin{subfigure}[a]{7.5cm}
    \includegraphics[width=7.5cm]{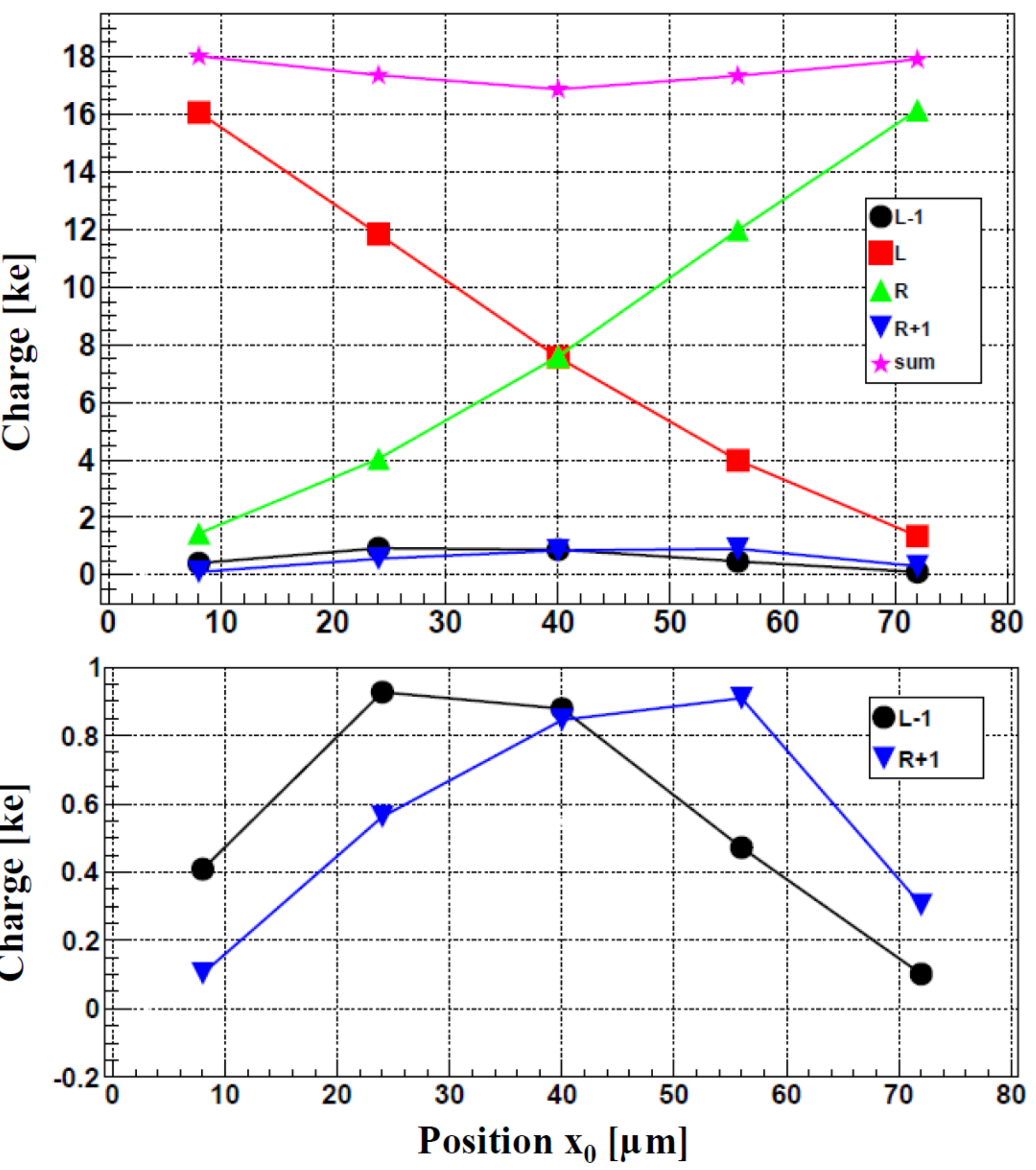}
    \caption{ }
     \label{fig:PHx2}
   \end{subfigure}%
   \caption{\,Distribution of the median values of the pulse heights in the individual readout strips and PH(4-cluster)  as functions of track position for the non-hadron-irradiated MCz $p$-stop sensor biased at 600\,V after a dose in the SiO$_2$ from the $\beta $\,source of
   (a)\,10\,Gy and (b)\,450\,Gy. The bottom plots show PH($L$-1) and PH($R$+1) with an expanded $y$\,scale.
   The statistical errors for the central bins ($x_0 = 40\,\upmu$m) are $20\,e$ for PH($L$-1) and PH($R$+1), $80\,e$ for PH($L$) and PH($R$), and $160\,e$ for the sum.}
  \label{fig:PHx}
\end{figure}

 As discussed in Sect.\,\ref{sect:Analysis}, the measured value of $\eta $ allows the estimation of the distance, $x_0$, of the particle passage from the centre of the readout strip $L$ and thus the investigation of the position dependence of the charge collection.
 Fig.\,\ref{fig:PHx} shows, for the non-hadron-irradiated $p$-stop sensor fabricated on magnetic-Czochralski silicon biased at 600\,V, the median values of PH(4-cluster), PH($L$-1), PH($L$), PH($R$), and PH($R$+1) as functions of $x_0$ for $\beta $\,source dose values of 10 and 450\,Gy.
 Comparing the distributions, it can be seen that for the 450\,Gy data in the region between the readout strips, PH(4-cluster) decreases by about 12\,\%, and the pulses induced in strips $L$-1 and $R$+1 increase from about 3\,\% to 5\,\%.
 In Sect.\,\ref{sect:Discussion} we will present a possible explanation of this observation.

 \section{Discussion of the results}
    \label{sect:Discussion}

 Figs.\,\ref{fig:Pstop} and \ref{fig:Pspray} show that significant changes in the charge collection are already observed after 0.2\,days, when the ionizing dose in the SiO$_2$ at the maximum of the distribution is 10\,Gy, and the NIEL is $2\times 10^{7}$\,n$_{eq}$/cm$^{2}$.
 Given such a low NIEL value, bulk damage is excluded as an explanation, and charge buildup in the insulators and surface damage have to be considered.

 In SiO$_2$ an average energy loss of $17 \pm 1$\,eV is required to produce an $eh$\,pair\,\cite{Ma:1989, Oldham:1999}, resulting in a density of generated $eh$\,pairs of $8.2\,\times 10^{14}$\,cm$^{-3}$\,Gy$^{-1}$.
 Thus, for a SiO$_2$ thickness of 1150\,nm, a dose of 10\,Gy generates $9.4 \times 10^{11} $cm$^{-2}\,eh$\,pairs.
 A fraction of the $eh$\,pairs will annihilate on a time scale of picoseconds.
 As shown in Fig.\,1.7 of Ref.\,\cite{Oldham:1999} and in Ref.\,\cite{Zhang:Thesis} this fraction depends on the ionization density and the local electric field:
 For 12\,MeV electrons it is about 70\,\% at zero field, and 10\,\% at 1\,MV/cm.
 The corresponding numbers for 10\,keV X-rays are 90\,\% and 45\,\%, respectively.
 Electrons, thanks to their high mobility (approximately 20\,cm$^2$/(V$\cdot $s) at room temperature) will leave the oxide.
 However holes, which have a mobility in the range $10^{-4}$ to $10^{-11}$\,cm$^2$/(V$\cdot $s) at room temperature and even lower values at $-20\,^\circ $\,C, will charge up the oxide.
 As the electric field in the SiO$_2$ points towards the Si, the holes will eventually reach the Si-SiO$_2$\,interface, where they either escape into the Si or are trapped as oxide charges.

 \begin{figure}[!ht]
  \centering
   \includegraphics[width=15cm]{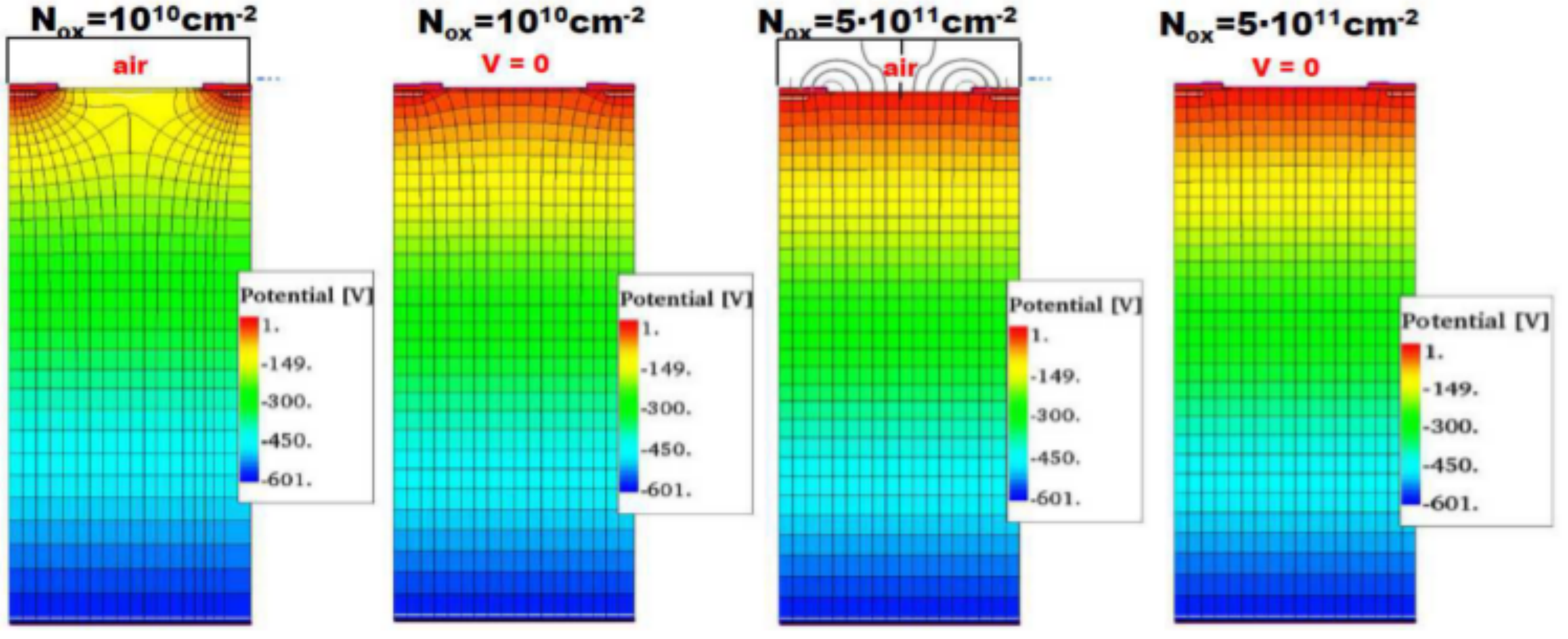}
    \caption{Simulated potential and electric field distributions for a $p$-spray sensor with an integral of the $p$-spray doping concentration of $2\times 10^{11}$\,cm$^{-2}$, and oxide-charge  densities of $10^{10}$ and $5 \times 10^{11}$ \,cm$^{-2}$.  The different surface boundary conditions are indicated in the figure and discussed in the text.}
   \label{fig:Efield}
 \end{figure}

 2-D simulations with SYNOPSYS TCAD\,\cite{SYNOPSYS} were  made to calculate the electric field in the sensor for different oxide-charge  densities, and boundary conditions on the sensor surface.
 The results are presented in Fig.\,\ref{fig:Efield}, which shows the electric field stream lines and the electric potential for a sensor of thickness 200\,$\upmu $m, bulk $p$\,doping of $3.7 \times 10^{12}$\,cm$^{-3}$ and a $p$-spray dopant concentration of $2\times 10^{11}$\,cm$^{-2}$, at a bias voltage of 600\,V.
 For the layout of the SiO$_2$ and Si$_3$N$_4$ layers, and of the $n^+$\,implants used in the simulations, we refer to Fig.\,\ref{fig:Sensor2}.
 For the two simulations on the left side an oxide-charge  density of $N_{ox} = 10^{10}$\,cm$^{-2}$ was assumed, and for the two simulations on the right side $N_{ox} = 5 \times 10^{11}$\,cm$^{-2}$.
 Two different boundary conditions at the strip side were investigated: A potential of 0\,V on a plane at 500\,$\upmu $m distance from the sensor surface, and zero charge density on the sensor surface (denoted \textit{air}), and a potential of 0\,V on the sensor surface (denoted \textit{V\,$ = 0$}).
 The first boundary condition corresponds to the situation shortly after applying the voltage to the sensor, when the charge distribution on the sensor surface is the same as before applying the voltage.
 The second one corresponds to the equilibrium situation of the biased sensor, when surface charges have moved, until the longitudinal electric field on the surface vanishes.
 The time constant for reaching the equilibrium is a strong function of humidity and temperature, and can be as long as several days\,\cite{Poehlsen:2013}.

 We note that for the simulation with $N_{ox} = 10^{10}$\,cm$^{-2}$, and in particular for the \textit{air} boundary condition, most field stream lines originate at the readout strips.
 Thus for a particle at normal incidence, practically all generated electrons will reach a single readout strip, and apart from a small effect due to charge diffusion, there will be no charge division.

 For the simulation with $N_{ox} = 5 \times 10^{11}$\,cm$^{-2}$ the field distribution is very different, resembling the field of a pad diode.
 The reason is that the positive oxide-charge density, $N_{ox}$, is bigger than the negative charge density of the $p$-spray doping, which results in an approximately constant potential at the Si-SiO$_2$ interface.
 Therefore the electric-field component parallel to the interface is small, and electrons,  which reach the Si-SiO$_2$ interface within the typical charge collection time of a few nanoseconds, will drift to the readout strips on a much longer time scale.
 If this time scale is long compared to the integration time of the readout electronics, the electrons are effectively trapped at the Si-SiO$_2$ interface, and signals will be recorded not only on strips $L$ and $R$, but also on strips $L$-1 and $R$+1, and even beyond.
 This will be discussed in more detail below.
 Electronic cross-talk will also produce signals in strips $L$-1 and $R$+1.
 Simulations of the electric field in an $n^+p$ sensor with $p$-stop isolation for oxide-charge densities of $10^{10}$ and $10^{12}$\,cm$^{-2}$ are also reported in Ref.\,\cite{Unno:2013}.
 The results agree with our simulations; however the effects on charge collection had not been studied.


 Fig.\,\ref{fig:PHx} shows that signals are actually observed in all four strips $L$-1, $L$, $R$ and $R$+1 for particles passing between two readout strips.
 We also find that when the dose changes from 10\,Gy to 450\,Gy, the signals in the next-to-next-neighbour strips increase, and the cluster pulse height, PH(4-cluster), decreases.
 As the electronic cross-talk is the same in both cases, we conclude that the radiation from the $\beta $\,source changes the electric field in the sensor, which causes the observed changes with radiation from the $\beta $\,source.

  \begin{figure}[!ht]
  \centering
   \includegraphics[width=12cm]{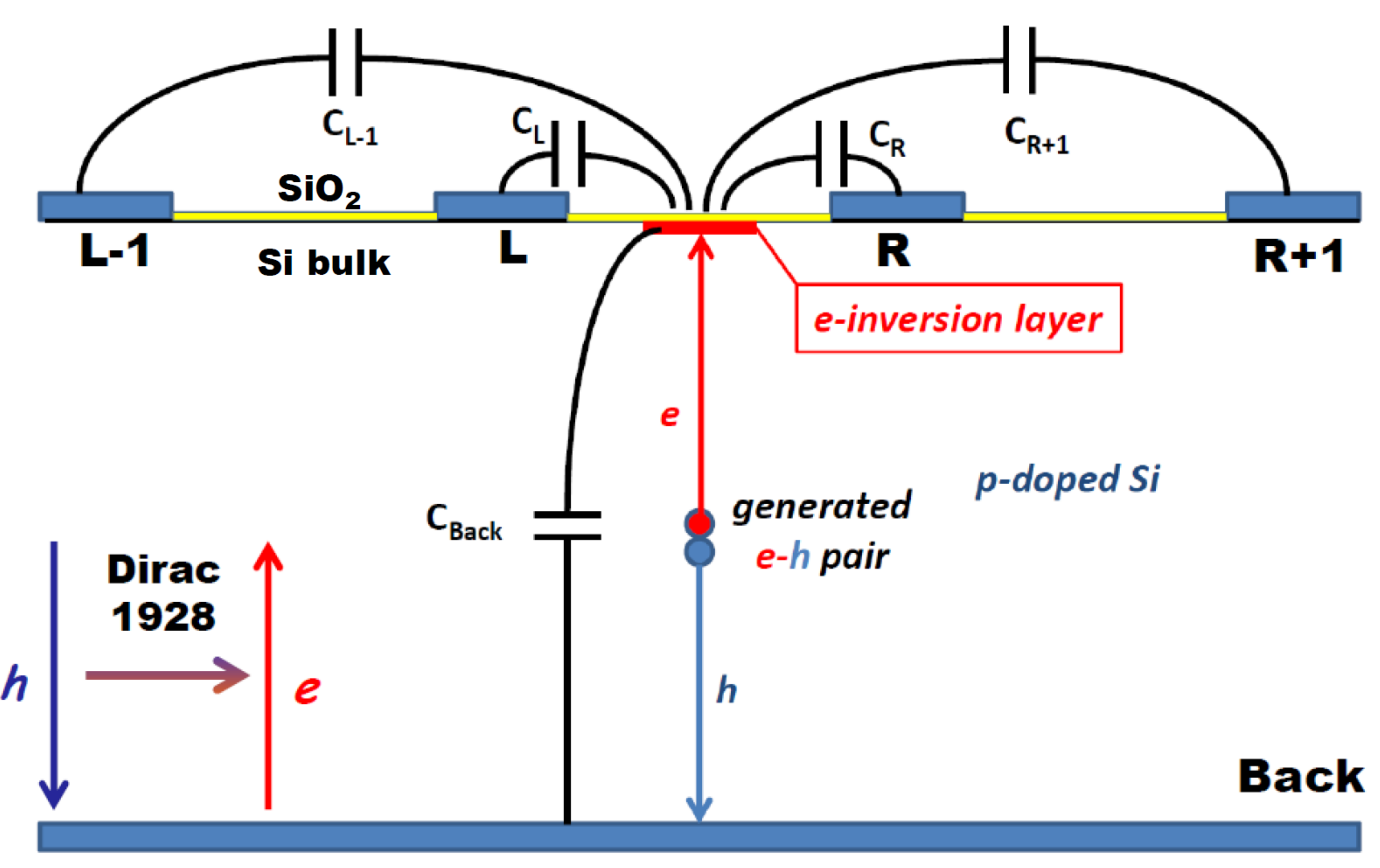}
    \caption{Capacitance model of a strip sensor used to discuss charge sharing and charge losses in segmented silicon sensors. The insert \textit{Dirac\,1928} is a reminder that an electron with momentum $\vec{\,p} $ is equivalent to its antiparticle with momentum $-\vec{\,p} $.}
   \label{fig:Capmodel}
 \end{figure}

 Signal formation in segmented sensors is usually described with the help of weighting fields\,\cite{Shockley:1938, Ramo:1939, Gunn:1964}, which describe the capacitative coupling of a charge at a given position in the sensor to the individual electrodes.
 For the following qualitative discussion we prefer the equivalent description in terms of effective coupling capacitors as presented in Ref.\,\cite{Koetz:1985} and sketched in Fig.\,\ref{fig:Capmodel}.
 It is more easily understood than the abstract weighting fields and allows an estimation of the cross-talk and the effect of the finite input capacitance of the readout electronics in a straightforward way.

 For a charge positioned at the Si-SiO$_2$ interface (e.g. at the head of the red arrow labeled $e$ in Fig.\,\ref{fig:Capmodel}) we label the coupling capacitors between the position of the electron and the individual strips, $C_i$, and the coupling capacitor to the backplane, $C_{Back}$.
 The figure is simplified, as it neither shows the capacitances between the readout strips, nor the capacitances of the $AC$ coupling nor the finite effective input impedance of cables and  readout electronics, which are responsible for cross-talk between readout channels.
 In the following discussion we assume that all electrodes are virtual grounds.
 This is a good approximation if the $AC$-coupling capacitances between the $n^+$\,implants and the Al electrodes are large and the effective input impedance of the readout is small compared to the inter-strip and strip-to-backplane capacitances.
 Including inter-strip capacitances, coupling capacitances and finite input impedances for the readout is straightforward, but not necessary for the following  discussion.

 We first consider the case of a single $eh$ pair generated somewhere in the sensor without charge trapping in the silicon crystal.
  In an $n^+p$\, sensor the holes drift to the back contact and the electrons towards the strip side, thereby inducing current transients in all electrodes.
 The charge induced in an electrode by a hole moving to the backplane, is equal to the charge induced by an electron moving from the backplane to the position where the hole was created.
 Thus, a hole from an $eh$\,pair moving to the backplane and the electron moving to the strip side, is equivalent to an electron moving all the way from the backplane to the strip side.
 As all electrodes are considered to be on virtual ground, an electron with charge $e$ that moves from the backplane to a readout strip induces in this strip and in the back electrode, a current transient whose integral is equal to $e$.
 Currents are also induced in all other strips, but their integrals are zero.
 However, an electron trapped at the Si-SiO$_2$ interface corresponds to charging up the capacitors $C_i$, which are connected in parallel to the virtual ground, such that signals $Q_i = e \times C_i/\Sigma \,C_j$ are induced in the electrodes $i$.
 The cluster charge, which is the sum of all signals induced in the readout strips, is $e \times (1-C_{Back}/\Sigma \,C_j)$, and the charge loss is $e \times C_{Back}/\Sigma \, C_j$.
 If more than one $eh$\,pair is produced, the contributions from all $eh$\,pairs have to be added.

 We can explain the increase of the charge losses with dose,
 seen at $x_0 = 40\,\upmu$m in Fig.\,\ref{fig:PHx}, with the formation of an inversion layer at the Si-SiO$_2$ interface:
 The capacitance to the back plane, $C_{Back}$, is approximately proportional to the width of the inversion layer, whereas the capacitances of the inversion layer to the strips $L$ and $R$ depend only weakly on its width\,\cite{Cattaneo:2010}.
 Thus, the wider the inversion layer, the larger the value of $C_{Back}/\Sigma \, C_j$, and the bigger the charge losses.
 The reason for the increase with dose of the charges induced in the strips $L$-1 and $R$+1, seen in the bottom of Fig.\,\ref{fig:PHx}, is not yet understood and needs further studies.

 In the case described, where charge trapping occurs only at the Si-SiO$_2$\,interface,
 the signals induced in the readout strips do not depend on the position where the $eh$\,pairs were generated, but only on the position at which the electrons arrive at the strip side.
 If charge trapping in the silicon bulk occurs, which is the case for silicon sensors with significant bulk damage by hadrons, the situation is significantly more complicated and better treated using weighting fields.
 However, the calculation of weighting fields in bulk-damaged silicon sensors is quite complicated\,\cite{Hamel:2008}, and its use for sensors with significant bulk damage and charge trapping is far from obvious.

 Hadron irradiation in silicon crystals produces defects with energy levels in the band gap.
 They cause generation currents and charge trapping.
 They also modify the electric field in the sensors\,\cite{Lindstroem:2003}.
 For hadron fluences above typically $10^{12}$\,n$_{eq}$/cm$^2$ the defect densities exceed the initial doping concentration, and radiation damage effects become significant.
 The generation current in an $n^+p$ sensor is dominated by holes at the back side and electrons at the strip side.
 Therefore, the fractions of occupied radiation-induced donors and acceptors vary with position, and the effective doping is strongly position dependent.
 Finally, the electric field in the sensor is the result of the interplay between the initial doping concentration, the additional effective doping from the charged radiation-induced defects, and the density of free charge carriers.
 We refer to Refs.\,\cite{Menichelli:1999, Eremin:2002}, where these effects have been discussed for the first time for silicon pad sensors.
 The high electron density in the region of the readout strips results in an electric field, which is qualitatively similar to the one shown in Fig.\,\ref{fig:Efield} left.
 This is our explanation for the low charge sharing and the reduced influence of the radiation-induced oxide-charge on the charge collection observed for the hadron-irradiated sensor and shown in Fig.\,\ref{fig:Irrad}.
  We note that the $CS$\,value for the hadron-irradiated sensor is close to the value expected from the angular distribution of the trigger electrons. 

 Next, we discuss the observation that the initial value and the initial dose-dependence of $CS$ for the non-hadron-irradiated $p$-stop and $p$-spray sensor are different, as can be seen in Figs.\,\ref{fig:Pstop2} and \ref{fig:Pspray2}.
 For the $p$-stop sensor the initial value is about 50\,\% and initially decreases with dose, whereas for the $p$-spray sensor it is about 30\,\% and increases.
 Our explanation is qualitative and should be confirmed using  TCAD simulations.
 However, there is the problem that the parameters, which describe the buildup of oxide charges for high fields at the Si-SiO$_2$ interface, are not known.

 For the $p$-spray sensor the negative space charge of the depleted $p$-spray implant shields the positive oxide charge, whilst for the $p$-stop sensor this shielding occurs only at the narrow $p$-stop implants.
 Therefore more field lines will originate at the Si-SiO$_2$ interface for the $p$-stop sensor and the initial charge sharing will be larger.
 Our explanation for the initial difference in the change of the charge sharing is more speculative.
 The initial decrease of the charge sharing for the $p$-stop sensor may be explained by the formation of a local electron inversion layer at the Si-SiO$_2$ interface below the metal overhang.
 There, the electric field in the SiO$_2$ is high and the $eh$\,recombination probability in the SiO$_2$ is low, which leads to an accelerated buildup of oxide charges.
 Such a local inversion layer effectively extends the charge-collection region of the readout strips and thus reduces the charge sharing.

 After about 4 days of irradiation with the $\beta $\,source, which corresponds to a dose of about 200\,Gy, $CS$ saturates and  the following values are reached: $\approx 80$\,\% for the $p$-stop and $\approx 70$\,\% for the $p$-spray sensor.
 We have not checked if the $CS$\,values become equal at higher surface doses, which one might naively expect.

  We expect that the surface damage has opposite effects on the breakdown voltage for $n^+p$ sensors and for $p^+n$ sensors.
 For $n^+p$ sensors, radiation-induced positive oxide charges compensate the negative charges of the $p$\,doping, thus reducing the electric field below the metal overhang and at the corners of the $n^+$\,implants.
 The breakdown voltage is therefore expected to increase with  increasing surface damage.
 For $p^+n$ sensors, the positive oxide charges add to the positive charges from the $n$\,doping, which increases the electric field below the metal overhang and at the corner of the $p^+$\,implants, leading to a decrease in breakdown voltage with increasing surface damage.
 Such a decrease has actually been observed and is documented in Refs.\,\cite{Erfle:2014, Ohsugi:1996, Schwandt:2013}.
 As shown in Refs.\,\cite{Schwandt:2014, Schwandt:Thesis}, a special optimization of the sensor design can avoid this problem.

 The annealing studies have only been made for the $p$-stop sensor and the results are shown in Fig.\,\ref{fig:Anneal}.
 Significant effects have been found at annealing temperatures of 60\,$^\circ $C and 80\,$^\circ $C.
 After annealing $CS$ increased by  $\approx 10$\,\%, but during the following period of approximately 7\,hours of irradiation with the $\beta $\,source, it decreased again by $\approx 20$\,\%.
 After the different annealing steps the long-term measurement showed that the asymptotic value of $CS$, which is $\approx 80$\,\%, is again approached after about 4 days.


  \section{Summary and outlook}
    \label{sect:Summary}

 The measurements presented in this publication show that an ionizing dose as low as 10\,Gy can cause significant changes in the charge collection and charge sharing of segmented $n^+p$ silicon sensors.
 For a sensor with a $p$-spray implant
 between the $n^+$\,implants of the readout electrodes it is found that, once the positive oxide-charge density resulting from the ionizing radiation in the oxide exceeds the negative charge density of the depleted $p$\,implants, the field in the sensor is similar to the field in a pad diode.
 In addition, electron-inversion layers can form at the Si-SiO$_2$ interface if the positive oxide-charge density is sufficiently high.
 Under these conditions some charge losses and approximately ideal charge division between the two readout strips adjacent to the particle passage were observed.
 The results for sensors with $p$-stop implants were qualitatively similar.
 As the radiation-induced positive oxide-charge density saturates for doses above $\approx 100$\,kGy at a value of a few $10^{12}$\,cm$^{-2}$\,\cite{Zhang:2012}, we expect that for $p$-spray dopant concentrations exceeding these values, no significant changes with dose in charge sharing and charge losses due to surface damage will occur.
 For a hadron-irradiated sensor the observed changes with ionizing dose were significantly smaller than for the non-hadron-irradiated sensors.
 Qualitative and sometimes speculative explanations are given, which are based on simplified TCAD simulations of the electric field in the sensor.
 For a quantitative understanding more detailed TCAD simulations will be required.
 However, precise information on surface radiation damage, the charging up of the insulators, high-field charge injection into insulators, and on the boundary conditions on the sensor surface, is presently not available.
 Last but not least, the study shows the importance of taking both bulk and surface damage into account when designing silicon sensors for low- and high-radiation environments.

 The impact of changes in charge sharing and charge losses on the performance of silicon detectors with high spatial resolution depends very much on the type and performance of the readout electronics used.
 For analogue readout with a good signal-to-noise-ratio, $S/N$, charge sharing can be used to optimize the spatial resolution, which is then given by the product of pitch and $N/S$.
 However, if the $S/N$ is poor and a relatively high readout threshold has to be set, it will be difficult to reach a high detection efficiency.
 As shown in this paper for a strip sensor with 80\,$\upmu $m pitch, the pulse height of the seed strip is frequently smaller than 50\,\% of the most probable value (mpv) of the cluster pulse-height distribution.
 To reach high efficiencies, the pulse-height threshold has to be below $0.4 \times $\,mpv.
 If a pulse-height threshold of 3 times the variance of the noise is required to limit the number of noise hits, a $S/N$ value for the cluster signal of at least 7.5 is required.
 If binary readout is used, the threshold requires a careful optimization with respect to noise hits, efficiency, cluster size and position resolution.
 We conclude that it is important to take charge sharing properly into account when optimizing segmented silicon sensors.

  \section*{Acknowledgements}
    \label{sect:Acknowledgements}

 The research leading to these results has received funding from the European Commission under the FP7 Research Infrastructures project AIDA, grant agreement no. 262025.
 The information herein only reflects the views of its authors and not those of the European Commission and no warranty expressed or implied is made with regard to such information or its use.

 The authors thank W.\,G\"artner and the workshop of the Institute for Experimental Physics for designing and constructing the $\beta $\,setup, and P.\,Buhmann, R.-P.\,Feller, M.\,Matysek and R.\,Mohrmann for the continuous improvement and maintenance of the measurement infrastructure of the detector laboratory as well as for helping with the measurements.
  We are also grateful to the HGF Alliance \emph{Physics at the Terascale}, which has partially  funded the setup used for the measurements, and to the BMBF, the Ministry of Research and Education of the German Federal Republic, for funding the PhD\,position of J.\,Erfle within the Forschungsschwerpunkt FSP-102, \emph{Elementarteilchenphysik mit dem CMS-Experiment}.

\newpage

\section{Appendix: Author list}

{\centering
W.~Adam, T.~Bergauer, M.~Dragicevic, M.~Friedl, R.~Fruehwirth, M.~Hoch, J.~Hrubec, M.~Krammer, W.~Treberspurg, W.~Waltenberger
\par}

{\centering
\textit{Institut f\"ur Hochenergiephysik der \"Osterreichischen Akademie der Wissenschaften (HEPHY), Vienna, Austria}
\par}

{\centering
S.~Alderweireldt, W.~Beaumont, X.~Janssen, S.~Luyckx, P.~Van Mechelen,  N.~Van Remortel, A.~Van Spilbeeck
\par}

{\centering\itshape
Universiteit Antwerpen, Belgium
\par}

{\centering
P.~Barria, C.~Caillol, B.~Clerbaux, G.~De Lentdecker, D.~Dobur, L.~Favart, A.~Grebenyuk, Th.~Lenzi, A.~L\'eonard, Th.~Maerschalk, A.~Mohammadi, L.~Perni\`e, A.~Randle-Conde, T.~Reis, T.~Seva, L.~Thomas, C.~Vander Velde, P.~Vanlaer,
J.~Wang, F.~Zenoni
\par}

{\centering
\textit{Brussels-ULB, Belgium}
\par}

{\centering
S.~Abu Zeid, F.~Blekman, I.~De Bruyn, J.~D'Hondt, N.~Daci, K.~Deroover, N.~Heracleous, J.~Keaveney, S.~Lowette, L.~Moreels, A.~Olbrechts, Q.~Python, S.~Tavernier, P.~Van Mulders, G.~Van Onsem, I.~Van Parijs, D.A.~Strom
\par}

{\centering
\textit{Brussels-VUB, Belgium}
\par}

{\centering
S.~Basegmez, G.~Bruno, R.~Castello, A.~Caudron, L.~Ceard, B.~De Callatay, C.~Delaere, T.~Du Pree, L.~Forthomme, A.~Giammanco, J.~Hollar, P.~Jez, D.~Michotte, C.~Nuttens, L.~Perrini, D.~Pagano, L.~Quertenmont, M.~ Selvaggi, M.~Vidal Marono
\par}

{\centering\itshape
CP3/IRMP - Universit\'e catholique de Louvain - Louvain-la-Neuve -- Belgium
\par}

{\centering
N.~Beliy, T.~Caebergs, E.~Daubie, G.H.~Hammad
\par}

{\centering\itshape
University of Mons, Belgium
\par}

{\centering
J.~H\"ark\"onen, T.~Lamp\'en, P.~-R.~Luukka, T.~M\"aenp\"a\"a, T.~Peltola, E.~Tuominen, E.~Tuovinen
\par}

{\centering
\textit{Helsinki Institute of Physics, Finland}
\par}

{\centering
P.~Eerola
\par}

{\centering
\textit{University of Helsinki and Helsinki Institute of Physics, Finland}
\par}

{\centering
T.~Tuuva
\par}

{\centering\itshape
Lappeenranta University of Technology, Lappeenranta, Finland
\par}

{\centering
G.~Beaulieu, G.~Boudoul, C.~Combaret, D.~Contardo, G.~Gallbit, N.~Lumb, H.~Mathez,  L.~Mirabito, S.~Perries, D.~Sabes, M.~Vander Donckt, P.~Verdier, S.~Viret, Y.~Zoccarato
\par}

{\centering
\textit{Universit\'e de Lyon, Universit\'e Claude Bernard Lyon 1, CNRS/IN2P3, Institut de Physique Nucl\'eaire de Lyon,
France}
\par}

{\centering
J.-L.~Agram, E.~Conte, J.-Ch.~Fontaine
\par}

{\centering
\textit{Groupe de Recherches en Physique des Hautes Energies, Universit\'e de Haute Alsace, Mulhouse, France}
\par}

{\centering
J.~Andrea, D.~Bloch, C.~Bonnin, J.-M.~Brom, E.~Chabert, L.~Charles, Ch.~Goetzmann, L.~Gross, J.~Hosselet, C.~Mathieu, M.~Richer, K.~Skovpen
\par}

{\centering
\textit{Institut Pluridisciplinaire Hubert Curien, Universit\'e de Strasbourg, IN2P3-CNRS, Strasbourg, France}
\par}

{\centering
C.~Autermann, M.~Edelhoff, H.~Esser, L.~Feld, W.~Karpinski, K.~Klein, M.~Lipinski, A.~Ostapchuk, G.~Pierschel, M.~Preuten, F.~Raupach, J.~Sammet, S.~Schael, G.~Schwering, B.~Wittmer, M.~Wlochal, V.~Zhukov
\par}

{\centering\itshape
I.~Physikalisches Institut, RWTH Aachen University, Germany
\par}

{\centering
C.~Pistone, G.~Fluegge, A.~Kuensken, M.~Geisler, O.~Pooth, A.~Stahl
\par}

{\centering
\textit{III.~Physikalisches Institut, RWTH Aachen University, Germany}
\par}

{\centering
N.~Bartosik, J.~Behr, A.~Burgmeier, L.~Calligaris, G.~Dolinska, G.~Eckerlin, D.~Eckstein, T.~Eichhorn, G.~Fluke, J.~Garay Garcia, A.~Gizhko, K.~Hansen, A.~Harb, J.~Hauk, A.~Kalogeropoulos, C.~Kleinwort, I.~Korol, W.~Lange, W.~Lohmann, R.~Mankel, H.~Maser, G.~Mittag, C.~Muhl, A.~Mussgiller, A.~Nayak, E.~Ntomari, H.~Perrey, D.~Pitzl, M.~Schroeder, C.~Seitz, S.~Spannagel, A.~Zuber
\par}

{\centering\itshape
DESY, Hamburg, Germany
\par}

{\centering
H.~Biskop, V.~Blobel, P.~Buhmann, M.~Centis-Vignali, A.-R.~Draeger, J.~Erfle, E.~Garutti, J.~Haller, Ch.~Henkel, M.~Hoffmann, A.~Junkes, R.~Klanner, T.~Lapsien, S.~M\"attig, M.~Matysek, A.~Perieanu, J.~Poehlsen, T.~Poehlsen, Ch.~Scharf, P.~Schleper, A.~Schmidt, J.~Schwandt, V.~Sola, G.~Steinbr\"uck, B.~Vormwald, J.~Wellhausen
\par}

{\centering
\textit{University of Hamburg, Germany}
\par}

{\centering
T.~Barvich, Ch.~Barth, F.~Boegelspacher, W.~De Boer, E.~Butz, M.~Casele, F.~Colombo,  A.~Dierlamm, R.~Eber, B.~ Freund, F.~Hartmann\footnote{ Also at CERN}, Th.~Hauth, S.~Heindl, K.-H.~Hoffmann, U.~Husemann, A.~Kornmeyer, S.~Mallows, Th.~Muller, A.~Nuernberg, M.~Printz, H.~J.~Simonis, P.~Steck, M.~Weber, Th.~Weiler
\par}

{\centering
\textit{Karlsruhe-IEKP, Germany}
\par}

{\centering
A.~Bhardwaj, A.~Kumar, A.~Kumar, K.~Ranjan
\par}

{\centering\itshape
Department of Physics and Astrophysics, University of Delhi, Delhu, India
\par}

{\centering
H.~Bakhshiansohl, H.~Behnamian, M.~Khakzad, M.~Naseri
\par}

{\centering
\textit{Institute for Research in Fundamental Sciences (IPM), Tehran, Iran}
\par}

{\centering
P.~Cariola, G.~De Robertis, L.~Fiore, M.~Franco, F.~Loddo, G.~Sala, L.~Silvestris
\par}

{\centering\itshape
INFN Bari, Italy
\par}

{\centering
D.~Creanza, M.~De Palma, G.~Maggi, S.~My, G.~Selvaggi
\par}

{\centering\itshape
INFN and Dipartimento Interateneo di Fisica, Bari, Italy
\par}

{\centering
S.~Albergo, G.~Cappello, M.~Chiorboli, S.~Costa, F.~Giordano, A.~Di Mattia, R.~Potenza, M.A.~Saizu\footnote{ Also at
Horia Hulubei National Institute of Physics and Nuclear Engineering (IFIN-HH), Bucharest, Romania}, A.~Tricomi, C.~Tuv\`e
\par}

{\centering\itshape
INFN and University of CATANIA, Italy
\par}

{\centering
G.~Barbagli, M.~Brianzi, R.~Ciaranfi, C.~Civinini, E.~Gallo, M.~Meschini, S.~Paoletti,
\par}

{\centering
G.~Sguazzoni
\par}

{\centering\itshape
INFN Firenze, Italy
\par}

{\centering
V.~Ciulli, R.~D'Alessandro, S.~Gonzi, V.~Gori, E.~Focardi, P.~Lenzi, E.~Scarlini, A.~Tropiano, L.~Viliani
\par}

{\centering
\textit{INFN and University of Firenze, Italy}
\par}

{\centering
F.~Ferro, E.~Robutti
\par}

{\centering\itshape
INFN Genova, Italy
\par}

{\centering
M.~Lo Vetere
\par}

{\centering\itshape
INFN and University of Genova, Italy
\par}

{\centering
S.~Gennai, S.~Malvezzi, D.~Menasce, L.~Moroni, D.~Pedrini
\par}

{\centering
\textit{INFN Milano-Bicocca, Italy}
\par}

{\centering
M.~Dinardo, S.~Fiorendi, R.A.~Manzoni
\par}

{\centering\itshape
INFN and Universita degli Studi di Milano-Bicocca, Italy
\par}

{\centering
P.~Azzi, N.~Bacchetta
\par}

{\centering\itshape
INFN Padova, Italy
\par}

{\centering
D.~Bisello, M.~Dall'Osso, T.~Dorigo, P.~Giubilato, N.~Pozzobon, M.~Tosi, A.~Zucchetta
\par}

{\centering
\textit{INFN and University of Padova, Italy}
\par}

{\centering
F.~De Canio, L.~Gaioni, M.~Manghisoni, B.~Nodari, V.~Re, G.~Traversi
\par}

{\centering\itshape
INFN Pavia and University of Bergamo, Italy
\par}

{\centering
D.~Comotti, L.~Ratti
\par}

{\centering\itshape
INFN Pavia and University of Pavia, Italy
\par}

{\centering
G.~M.~Bilei, L.~Bissi, B.~Checcucci, D.~Magalotti\footnote{ Also at Modena and Reggio Emilia University, Italy}, M.~Menichelli, A.~Saha, L.~Servoli, L.~Storchi
\par}

{\centering\itshape
INFN Perugia, Italy
\par}

{\centering
M.~Biasini, E.~Conti, D.~Ciangottini, L.~Fan\`o, P.~Lariccia, G.~Mantovani, D.~Passeri, P.~Placidi, M.~Salvatore, A.~Santocchia, L.A.~Solestizi, A.~Spiezia
\par}

{\centering\itshape
INFN and University of Perugia, Italy
\par}

{\centering
N.~Demaria, A.~Rivetti
\par}

{\centering\itshape
INFN Torino, Italy
\par}

{\centering
R.~Bellan, S.~Casasso, M.~Costa, R.~Covarelli, E.~Migliore, E.~Monteil, M.~Musich, L.~Pacher, F.~Ravera, A.~Romero, A.~Solano, P.~Trapani
\par}

{\centering\itshape
INFN and University of Torino, Italy
\par}

{\centering
R.~Jaramillo Echeverria, M.~Fernandez, G.~Gomez, D.~Moya, F.J.~Gonzalez Sanchez, F.J.~Munoz Sanchez, I.~Vila, A.L.~Virto
\par}

{\centering
\textit{Instituto de F{\i}sica de Cantabria (IFCA), CSIC-Universidad de Cantabria, Santander, Spain}
\par}

{\centering
D.~Abbaneo,~I.~Ahmed,~E.~Albert,~G.~Auzinger,~G.~Berruti,~G.~Bianchi,~G.~Blanchot,
\par}

{\centering
H.~Breuker, D.~Ceresa, J.~Christiansen, K.~Cichy, J.~Daguin, M.~D'Alfonso, A.~D'Auria, S.~Detraz, S.~De Visscher, D.~Deyrail, F.~Faccio, D.~Felici,~N.~Frank,~K.~Gill, D.~Giordano, P.~Harris, A.~Honma, J.~Kaplon, A.~Kornmayer, M.~Kortelainen, L.~Kottelat, M.~Kovacs, M.~Mannelli, A.~Marchioro, S.~Marconi, S.~Martina, S.~Mersi, S.~Michelis, M.~Moll, A.~Onnela, T.~Pakulski, S.~Pavis, A.~Peisert, J.-F.~Pernot, P.~Petagna, G.~Petrucciani, H.~Postema, P.~Rose, M.~Rzonca, M.~Stoye, P.~Tropea, J.~Troska, A.~Tsirou, F.~Vasey, P.~Vichoudis, B.~Verlaat, L.~Zwalinski
\par}

{\centering
\textit{European Organization for Nuclear Research (CERN), Geneva, Switzerland}
\par}

{\centering
F.~Bachmair, R.~Becker, L.~B\"ani, D.~di Calafiori, B.~Casal, L.~Djambazov, M.~Donega, M.~D\"unser, P.~Eller, C.~Grab, D.~Hits, U.~Horisberger, J.~Hoss, G.~Kasieczka, W.~Lustermann, B.~Mangano, M.~Marionneau, P.~Martinez Ruiz del Arbol, M.~Masciovecchio, L.~Perrozzi, U.~Roeser, M.~Rossini, A.~Starodumov, M.~Takahashi, R.~Wallny
\par}

{\centering\itshape
ETH Z\"urich, Z\"urich, Switzerland
\par}

{\centering
C.~Amsler\footnote{ Now at University of Bern, Switzerland}, K.~B\"osiger, L.~Caminada, F.~Canelli, V.~Chiochia, A.~de Cosa, C.~Galloni, T.~Hreus, B.~Kilminster, C.~Lange, R.~Maier, J.~Ngadiuba, D.~Pinna, P.~Robmann, S.~Taroni, Y.~Yang
\par}

{\centering
\textit{University of Z\"urich, Switzerland}
\par}

{\centering
W.~Bertl, K.~Deiters, W.~Erdmann, R.~Horisberger, H.-C.~Kaestli, D.~Kotlinski, U.~Langenegger, B.~Meier, T.~Rohe, S.~Streuli
\par}

{\centering
\textit{Paul Scherrer Institut, Villigen, Switzerland}
\par}

{\centering
P.-H.~Chen, C.~Dietz, U.~Grundler, W.-S.~Hou, R.-S.~Lu, M.~Moya, R.~Wilken
\par}

{\centering
\textit{National Taiwan University, Taiwan, ROC}
\par}

{\centering
D.~Cussans, H.~Flacher, J.~Goldstein, M.~Grimes, J.~Jacob, S.~Seif El Nasr-Storey
\par}

{\centering\itshape
University of Bristol, Bristol, United Kingdom
\par}

{\centering
J.~Cole, P.~Hobson, D.~Leggat, I.~D.~Reid, L.~Teodorescu
\par}

{\centering
\textit{Brunel University, Uxbridge, United Kingdom}
\par}

{\centering
R.~Bainbridge, P.~Dauncey, J.~Fulcher, G.~Hall, A.-M.~Magnan, M.~Pesaresi, D.M.~Raymond, K.~Uchida
\par}

{\centering
\textit{Imperial College, London, United Kingdom}
\par}

{\centering
J.A.~Coughlan, K.~Harder, J.~Ilic, I.R.~Tomalin
\par}

{\centering
\textit{STFC, Rutherford Appleton Laboratory, Chilton, Didcot, United Kingdom}
\par}

{\centering
A.~Garabedian, U.~Heintz, M.~Narain, J.~Nelson, S.~Sagir, T.~Speer, J.~Swanson, D.~Tersegno, J.~Watson-Daniels
\par}

{\centering\itshape
Brown University, Providence, Rhode Island, USA
\par}

{\centering
M.~Chertok, J.~Conway, R.~Conway, C.~Flores, R.~Lander, D.~Pellett, F.~Ricci-Tam,M.~Squires, J.~Thomson, R.~Yohay
\par}

{\centering
\textit{University of California, Davis, California, USA}
\par}

{\centering
K.~Burt, J.~Ellison, G.~Hanson, M.~Malberti, M.~Olmedo
\par}

{\centering\itshape
University of California, Riverside, California, USA
\par}

{\centering
G.~Cerati, V.~Sharma, A.~Vartak, A.~Yagil, G.~Zevi Della Porta
\par}

{\centering
\textit{University of California, San Diego, California, USA}
\par}

{\centering
V.~Dutta, L.~Gouskos, J.~Incandela, S.~Kyre, N.~McColl, S.~Mullin, D.~White
\par}

{\centering
\textit{University of California, Santa Barbara, California, USA}
\par}

{\centering
J.~P.~Cumalat, W.~T.~Ford, A.~Gaz, M.~Krohn, K.~Stenson, S.R.~Wagner
\par}

{\centering
\textit{University of Colorado, Boulder, Colorado, USA}
\par}

{\centering
B.~Baldin, G.~Bolla, K.~Burkett, J.~Butler, H.~Cheung, J.~Chramowicz, D.~Christian, W.E.~Cooper, G.~Deptuch, G.~Derylo, C.~Gingu, S.~Gruenendahl, S.~Hasegawa, J.~Hoff, J.~Howell, M.~Hrycyk, S.~Jindariani, M.~Johnson, A.~ Jung, U.~Joshi, F.~Kahlid, C.~M.~Lei, R.~Lipton, T.~Liu, S.~Los, M.~Matulik, P.~Merkel, S.~Nahn, A.~Prosser, R.~Rivera, A.~Shenai, L.~Spiegel, N.~Tran, L.~Uplegger, E.~Voirin, H.~Yin
\par}

{\centering
\textit{Fermi National Accelerator Laboratory (FNAL), Batavia, Illinois, USA}
\par}

{\centering
M.R.~Adams, D.R.~Berry, A.~Evdokimov, O.~Evdokimov, C.E.~Gerber, D.J.~Hofman, B.K.~Kapustka, C.~O'Brien, D.I.~Sandoval Gonzalez, H.~Trauger, P.~Turner
\par}

{\centering
\textit{University of Illinois, Chicago, Illinois, USA}
\par}

{\centering
N.~Parashar, J.~Stupak, III
\par}

{\centering\itshape
Purdue University Calumet, Hammond, Indiana, USA
\par}

{\centering
D.~Bortoletto, M.~Bubna, N.~Hinton, M.~Jones, D.H.~Miller, X.~Shi
\par}

{\centering\itshape
Purdue University, West Lafayette, Indiana, USA
\par}

{\centering
P.~Tan
\par}

{\centering
\textit{University of Iowa, Iowa City, Iowa, USA}
\par}

{\centering
P.~Baringer, A.~Bean, G.~Benelli, J.~Gray, D.~Majumder, D.~Noonan, S.~Sanders, R.~Stringer
\par}

{\centering\itshape
University of Kansas, Lawrence, Kansas, USA
\par}

{\centering
A.~Ivanov, M.~Makouski, N.~Skhirtladze, R.~Taylor
\par}

{\centering\itshape
Kansas State University, Manhattan, Kansas, USA
\par}

{\centering
I.~Anderson, D.~Fehling, A.~Gritsan, P.~Maksimovic, C.~Martin, K.~Nash, M.~Osherson, M.~Swartz, M.~Xiao
\par}

{\centering\itshape
Johns Hopkins University, Baltimore, Maryland, USA
\par}

{\centering\itshape
Massachusetts Institute of Technology, Cambridge, Massachusetts, USA
\par}

{\centering
J.G.~Acosta, L.M.~Cremaldi, S.~Oliveros, L.~Perera, D.~Summers
\par}

{\centering\itshape
University of Mississippi, Mississippi, USA
\par}

{\centering
K.~Bloom, S.~Bose, D.R.~Claes, A.~Dominguez, C.~Fangmeier, R.~Gonzalez Suarez, F.~Meier, J.~Monroy
\par}

{\centering\itshape
University of Nebraska, Lincoln, Nebraska, USA
\par}

{\centering
K.~Hahn, S.~Sevova, K.~Sung, M.~Trovato
\par}

{\centering\itshape
Northwestern University, Evanston, Illinois, USA
\par}

{\centering
E.~Bartz, D.~Duggan, E.~Halkiadakis, A.~Lath, M.~Park, S.~Schnetzer, R.~Stone, M.~Walker
\par}

{\centering\itshape
Rutgers University, Piscataway, New Jersey, USA
\par}

{\centering
S.~Malik, H.~Mendez, J.E.~Ramirez Vargas
\par}

{\centering\itshape
University of Puerto Rico, Mayaguez, Puerto Rico, USA
\par}

{\centering
M.~Alyari, J.~Dolen, J.~George, A.~Godshalk, I.~Iashvili, J.~Kaisen, A.~Kharchilava, A.~Kumar, S.~Rappoccio
\par}

{\centering\itshape
State University of New York, Buffalo, New York, USA
\par}

{\centering
J.~Alexander, J.~Chaves, J.~Chu, S.~Dittmer, G.~Kaufman, N.~Mirman, A.~Ryd, E.~Salvati, L.~Skinnari, J.~Thom, J.~Thompson, J.~Tucker, L.~Winstrom
\par}

{\centering\itshape
Cornell University, Ithaca, New York, USA
\par}

{\centering
B.~Akg\"un, K.M.~Ecklund, T.~Nussbaum, J.~Zabel
\par}

{\centering\itshape
Rice University, Houston, Texas, USA
\par}

{\centering
B.~Betchart, R.~Covarelli, R.~Demina, O.~Hindrichs, G.~Petrillo
\par}

{\centering\itshape
University of Rochester, New York, USA
\par}

{\centering
R.~Eusebi, I.~Osipenkov, A.~Perloff, K.A.~Ulmer
\par}

{\centering\itshape
Texas A\&M University, College Station, Texas, USA
\par}

{\centering
A.~G.~Delannoy, P.~D'Angelo, W.~Johns
\par}

{\centering\itshape
Vanderbilt University, Nashville, Tennessee, USA
\par}

 \newpage

\end{document}